\documentclass[11pt]{article}
\pdfoutput=1

\usepackage{amsmath}
\usepackage{amssymb}
\usepackage{amsfonts}
\usepackage{mathrsfs}

\usepackage{mathrsfs}
\usepackage{fullpage}
\usepackage{setspace}
\usepackage{bm}
\usepackage{bbm}
\usepackage{relsize}
\usepackage{adjustbox}
\usepackage{multirow}
\usepackage{cite}
\usepackage{filecontents}
\usepackage{scalerel}

\usepackage[normalem]{ulem}
\usepackage{enumerate}
\usepackage{yfonts}

\usepackage{psfrag}
\usepackage{array}
\usepackage{booktabs}
\newcolumntype{N}{>{\centering\arraybackslash}m{.5in}}
\newcolumntype{G}{>{\centering\arraybackslash}m{2in}}

\usepackage[latin1]{inputenc}
\usepackage{graphicx}
\usepackage{cancel}
\usepackage{slashed}
\usepackage{mathtools}

\usepackage[font=small,labelfont=bf]{caption}
\usepackage{subcaption}

\usepackage[colorlinks=true]{hyperref} 
\hypersetup{
    unicode=false,          
    pdftoolbar=true,        
    pdfmenubar=true,        
    pdffitwindow=false,     
    pdfstartview={FitH},    
    pdftitle={Prospects of detecting deviations to Kerr geometry with EMRIs},    
    pdfauthor={Mostafizur Rahman, Shailesh Kumar and Arpan Bhattacharyya},     
    pdfnewwindow=true,      
    colorlinks=true,       
    linkcolor=blue,          
    citecolor=red,        
    filecolor=magenta,      
    urlcolor=blue           
    }

\usepackage{hyperref}
\hypersetup{colorlinks=true}

\def\equationautorefname~#1\null{%
	Eq.~(#1)\null
}
\def\figureautorefname~#1\null{%
	Fig.~#1\null
}
\def\tableautorefname~#1\null{%
	Table.~#1\null
}
\def\sectionautorefname~#1\null{%
	Section #1\null
}
\def\appendixautorefname~#1\null{%
	Appendix #1\null
}

\begin{document}

\numberwithin{equation}{section}
{
\begin{titlepage}
\begin{center}

\hfill \\
\hfill \\
\vskip 0.75in
{\Large \bf Prospects of detecting deviations to Kerr geometry with radiation reaction effects in EMRIs 
}\\

\vskip 0.3in

{\large
Abhishek Chowdhuri${}$\footnote{\href{mailto: chowdhuri_abhishek@iitgn.ac.in}{chowdhuri\_abhishek@iitgn.ac.in}}, Arpan Bhattacharyya${}$\footnote{\href{mailto:  abhattacharyya@iitgn.ac.in}{abhattacharyya@iitgn.ac.in}}} and Shailesh Kumar${}$\footnote{\href{mailto: shailesh.k@iitgn.ac.in}{shailesh.k@iitgn.ac.in}}

\vskip 0.3in

{\it ${}$Indian Institute of Technology, Gandhinagar, Gujarat-382355, India}

\vskip.5mm

\end{center}

\vskip 0.35in

\begin{center} 
{\bf ABSTRACT }
\end{center}
Direct detection of gravitational waves and binary black hole mergers have proven to be remarkable investigations of general relativity. In order to have a definitive answer as to whether the black hole spacetime under test is the Kerr or non-Kerr, one requires accurate mapping of the metric. Since EMRIs are perfect candidates for space-based detectors, Laser Interferometer Space Antenna (LISA) observations will serve a crucial purpose in mapping the spacetime metric. In this article, we consider such a study with the Johannsen spacetime that captures the deviations from the Kerr black hole and further discuss their detection prospects. We analytically derive the leading order post-Newtonian corrections in the average loss of energy and angular momentum fluxes generated by a stellar-mass object exhibiting eccentric equatorial motion in the Johannsen background. We further study the orbital evolution of the inspiralling object within the adiabatic approximation. We lastly provide the possible detectability of deviations from the Kerr black hole by estimating gravitational wave dephasing and highlight the crucial role of LISA observations. 
\vfill


\end{titlepage}
}

\newpage
\tableofcontents

\section{Introduction}
The series of detections by LIGO and Virgo collaborations \cite{LIGOScientific:2016aoc, LIGOScientific:2016vlm, LIGOScientific:2016sjg, LIGOScientific:2017bnn, LIGOScientific:2018mvr, LIGOScientific:2020ibl, LIGOScientific:2021djp} for gravitational waves (GWs) has opened up opportunities for new insights into the cosmos. GW astronomy is now a critical tool for answering longstanding non-trivial questions on astronomy and cosmology and offers a test of gravity in a strong and weak regime. The detectors boast a high level of precision, which will only improve in the years to come, thus demanding commensurately accurate theoretical predictions encoded in waveform templates, which will be utilized for detection and parameter estimations. These
waveforms are constructed from various techniques, including the
effective one-body (EOB) formalism\cite{Buonanno:1998gg, Buonanno:2000ef}, numerical relativity\cite{Pretorius:2005gq, Campanelli:2005dd, Baker:2005vv}, self-force techniques\cite{Mino:1996nk, Quinn:1996am}, and several perturbative methods for the inspiral phase, including the
post-Newtonian (PN)\cite{7c44806c-bf75-3a8d-b622-7761a1e00cd9, Blanchet:2013haa} and post-Minkowskian (PM) approximations\cite{Damour:2016gwp, Damour:2017zjx}, as well
as effective field theory (EFT) formalisms\cite{Porto:2016pyg, Schafer:2018kuf, Barack:2018yvs, Barack:2018yly, Levi:2018nxp}. Further improving these high-precision theoretical predictions from general relativity (GR) or other alternate gravity theories (non-GR) will be crucial, given the expected sensitivity improvements in the detectors.
\par
Here, we will focus on one of the above techniques, which has time and again proved to be a useful theoretical tool for building waveforms: the PN scheme. During the early inspiral phase, where the gravitational system is weak, the constituents of the binary black hole systems can be treated as non-relativistic, and one can impose an expansion in velocity squared which is of the order of inverse separation in
units of the Schwarzschild radius due to the virial theorem. This expansion has a well-established framework with a long history dating back to the leading 1PN correction to
the Newtonian gravitational potential \cite{7c44806c-bf75-3a8d-b622-7761a1e00cd9} with follow-up works on 2PN\cite{10.1143/PTP.50.492}, 3PN\cite{Jaranowski:1997ky, Damour:1999cr, Blanchet:2000nv, Damour:2001bu}, and 4PN \cite{Damour:2014jta, Jaranowski:2015lha, Bernard:2015njp, Marchand:2017pir, Foffa:2012rn, Foffa:2016rgu, Foffa:2019rdf, Porto:2017dgs, Foffa:2019yfl}
expressions for the conservative potential. More recently, 5PN static contributions have
also been computed\cite{Foffa:2019hrb, Blumlein:2019zku}. Apart from such analyses happening for GR, much work has been done in extending this PN scheme for non-GR theories as well\cite{Bernard:2022noq, Zhang:2017srh, Quartin:2023tpl, PhysRevD.106.064046}. The formalism can also be extended for hyperbolic encounters\cite{DeVittori:2014psa, Garcia-Bellido:2017knh, DeVittori:2012da}, for binaries moving in media \cite{Dai:2021olt, AbhishekChowdhuri:2023cle}, in the context of EFTs using the worldline EFT approaches\cite{Levi:2011eq, Levi:2015ixa, Maia:2017gxn, Mandal:2022ufb, Bhattacharyya:2023kbh, Diedrichs:2023foj}. In this paper,  we will only focus on the leading order term in this expansion, giving us the leading order contribution in the energy flux for binaries.  \par
While it is sufficient for weak-field regime analysis to rely on the PN approach, a model-independent strong-field test requires an underlying spacetime geometry. The most obvious one that is used is a deformed Kerr metric, with the deviations being parametric. Several
such frameworks have been suggested within
which possible observational signatures of a Kerr-like
black hole can be explored\cite{Collins:2004ex, Glampedakis:2005cf, PhysRevD.81.024030,PhysRevD.83.104027}.
\textcolor{black}{As these deviations are parametric, it follows that the observables obtained for these metrics should depend on one or more of these parameters.}. They have a smooth Kerr limit when the deviations vanish. These metrics can prove to be useful in validating GR. \textcolor{black}{Further, in view of testing the ``no-hair" theorems, these theories along with deviations have been studied in the strong field regime \cite{Psaltis:2008bb}, with the obvious expectation that it aligns with the weak field tests \cite{Will:2005va}.}
A deviation from the Kerr metric puts forward two possible interpretations. Within GR, the object, described by this kind of geometry, cannot be a black hole but
instead can be a stable stellar or an exotic
object \cite{Collins:2004ex, PhysRevD.87.124017}. 
\par 
However, such deformations do introduce some pathologies as well\cite{PhysRevD.87.124017}. In GR, the ``no-hair" theorem guarantees the ``Kerr Metric" is the only possible vacuum solution with two independent parameters: mass and spin. So, any deviation from this geometry will naturally violate the assumptions of this theorem, and this is exactly the reason that it might lead to situations such as the existence of closed time-like curves, null singularities, etc., hampering useful analysis in the strong field regime. In \cite{PhysRevD.87.124017}, authors analyze such parametric deviations in detail and explicitly show how they differ from the Kerr metric. Identifying the regions where these parameters are unphysical and the range of coordinates and parameters for which each extension remains regular is a possible way to circumvent these pathologies. Such a dependence is enough to test the no-hair theorem as long as these deviations are small. Large deviations, in turn, require understanding the strong-field radiative dynamics in the theory. To understand strong-field physics, one generally does not require the dynamical properties of the gravity theory and the underlying field equations. In most of the tests of strong field regime, one can generally find out that tests of the no-hair theorem in the electromagnetic spectrum\cite{Johannsen_2010, Johannsen1_2010, Johannsen_2011, PhysRevD.83.124015, Bambi:2014qna, Tsukamoto:2014tja, Bambi:2011jq, 2012ApJ...745....1P}, however, they are not a priori limited to the study of
small deviations from the Kerr metric because these are performed in a stationary black hole spacetime, where the metric serves as a fixed background.
\par
Several model-independent strong-field tests of the no-hair theorem have been suggested using GW observations of extreme mass ratio inspirals (EMRIs) \cite{Ryan:1995xi, PhysRevD.77.064022, PhysRevD.69.082005, PhysRevD.78.102002, PhysRevD.84.064016}, where an object (secondary with mass $\mu$) inspirals the central supermassive black hole (primary with mass $M$). Hence, exploring waveforms in this context can be a useful tool. As the name suggests, such binaries have a mass ratio lying in the range ($q\equiv\mu/M = 10^{-7}-10^{-4}$) and radiate GWs in the millihertz range, making it one of the most sought-after objects in LISA \cite{2017arXiv170200786A}. Much work has been done to look for theoretically modelling such signals \cite{Babak:2017tow, Amaro-Seoane:2007osp, Hinderer:2008dm, Drasco:2005kz, Rahman:2022fay, Rahman:2021eay, Maselli:2021men, Gair:2017ynp, LISA:2022kgy, Drummond:2023wqc, Fransen:2022jtw, PhysRevLett.125.221602, Bianchi:2020bxa} in the hope that they will infer the dynamics of the compact object deep
in the gravitational field of the central black hole. It allows us to probe intergalactic astrophysical environment \cite{Rahman:2023sof} and also place constraints for both modelling and data analysis of EMRIs. 
\par 
\textcolor{black}{Focusing our attention to the part of the parameter space where we can evade pathologies, we examine the effects of gravitational radiation reaction on orbits moving in the deformed Kerr geometry.}  In \cite{PhysRevD.104.064023}, these deformations in the context of Manko-Novikov type metrics have been studied to understand resonant glitches. We do our analysis by keeping only the leading order contribution coming due to the deviation parameters which enter in the expressions for various fluxes at 2PN order and to the leading order in mass ratio. We also assume adiabatic approximation, meaning the radiation reaction timescale is much longer than the orbital period; thus, the orbit looks geodesic on shorter timescales. The analysis we follow is similar to the one in \cite{Flanagan:2007tv, Ryan:1995xi}, where we use the leading order radiation reaction acceleration that acts on the particle. We keep the parameter responsible for axisymmetry breaking to analyse the effects it has on the phasing and dephasing of the gravitational waves\footnote{\textcolor{black}{This analysis is much different from the one in \cite{Fransen:2022jtw} where they have analysed scenarios for equatorial symmetry breaking for Kerr metric and their associated analytic Kludge waveforms}.}. In the adiabatic limit, the time-averaged rates governing the change in the constants of motion help us infer the orbital evolution. 
\par 
Before concluding, let us go through how we organize the paper: After a brief introduction on deformed geometries in Section (\ref{BH}), we discuss the geodesic motions of the particle in such geometry in Section (\ref{gdscvecmtn}). This gives us an idea of the orbital dynamics of the inspiralling object. Next, in Section (\ref{radiation reaction}), we introduce the radiation reaction formalism and discuss the changes in the orbital parameters. Section (\ref{dephasing1}) is solely focused on exploring the detection possibilities for our setup, where we go on to calculate dephasing. We conclude by discussing our results in Section (\ref{dscn}) with possible future directions. Some additional details are given in the appendices (\ref{apenteu1}) and (\ref{apenteu2}).\\
\par 
\textit{Notation and Convention: } We set the fundamental constants $G$ and $c$ to unity and  adopt positive sign convention $(-1,1,1,1)$. Roman letters are used to denote spatial indices, and Greek letters are used to represent four-dimensional indices. \newpage
\section{Deformed Kerr geometry: An introduction}\label{BH}
 We begin by considering the usual Kerr metric describing a rotating black hole spacetime. The line element is well-known and given by
\begin{align}
ds^{2} =& -\Sigma\frac{\Delta-a^{2}\sin^{2}\theta}{N}dt^{2}+\frac{\Sigma}{\Delta}dr^{2}+\Sigma d\theta^{2}-2a\Sigma\sin^{2}\theta \frac{(r^{2}+a^{2})-\Delta}{N}dt d\phi + \\ \nonumber 
& \Sigma \sin^{2}\theta\frac{(r^{2}+a^{2})^{2}-a^{2}\Delta\sin^{2}\theta}{N}d\phi^{2}\,,
\end{align}
where $\Sigma \equiv r^{2}+a^{2}\cos^{2}\theta$, $\Delta \equiv r^{2}-2Mr+a^{2}$ and $N \equiv \Sigma^{2}$. The above metric is Petrov Type D and admits a third constant of motion, the Carter constant $\mathcal{Q}$ apart from the energy $E$, and the axial angular momentum $J_{z}$. The Carter constant $\mathcal{Q}$ is then introduced while solving the Hamilton-Jacobi equation by the standard method, which assumes that it is separable. The equation reads like 
\begin{align}
g^{\alpha \beta}\frac{\partial}{\partial x^{\alpha}}\frac{\partial}{\partial x^{\beta}} = -\frac{1}{\Delta \Sigma}[-(r^{2}+a^{2})E+aJ_{z}]^{2}+\frac{1}{\Sigma \sin^{2}\theta}[J_{z}-aE\sin^{2}\theta]^{2}+\frac{\Delta}{\Sigma}\Big( \frac{\partial S_{r}}{\partial r}\Big)^{2}+\frac{1}{\Sigma}\Big( \frac{\partial S_{\theta}}{\partial \theta}\Big)^{2} \label{eq1}
\end{align}
where $S_{r}$ and $S_{\theta}$ are the radial and angular parts of the Hamilton-Jacobi function; see Eq.(\ref{ac1}). Upon rearranging the terms, we can write down two separate equations for the radial and angular parts and, in principle, solve them. However, Johannsen in \cite{Johannsen:2013szh} explored the idea of introducing deviations to Kerr metric such that the Hamilton-Jacobi still admits separability. After introducing scalar functions like $f(r)$, $g(r)$, $A_{i}(r)$ with $i=1, 2, 5$, and $A_{j}(\theta)$ with $j=3, 4, 6$, the Hamilton-Jacobi equation takes the following form:
\begin{align}
\begin{split}
g^{\alpha\beta}\frac{\partial}{\partial x^{\alpha}}\frac{\partial}{\partial x^{\beta}}=&-\frac{1}{\Delta \tilde{\Sigma}}\Big[(r^{2}+a^{2})A_{1}(r)\frac{\partial}{\partial t}+aA_{2}(r)\frac{\partial}{\partial \phi}\Big]^{2}+\frac{1}{\tilde{\Sigma} \sin^{2}\theta}\Big[A_{3}(\theta)\frac{\partial}{\partial \phi}+aA_{4}(\theta)\sin^{2}\theta\frac{\partial}{\partial t}\Big]^{2}\\ &+\frac{\Delta}{\tilde{\Sigma}}A_{5}(r)\Big( \frac{\partial}{\partial r}\Big)^{2}+\frac{1}{\tilde{\Sigma}}A_{6}(\theta)\Big( \frac{\partial}{\partial \theta}\Big)^{2}
\end{split}
\end{align}
where $\Tilde{\Sigma}=\Sigma+f(r)+g(\theta)$ \cite{Johannsen:2013szh}. Then similar to the Kerr case, one can come up with an equivalent expression to Eq.(\ref{eq1}):
\begin{align}\label{m1}
\begin{split}
g^{\alpha \beta}\frac{\partial}{\partial x^{\alpha}}\frac{\partial}{\partial x^{\beta}}=&-\frac{1}{\Delta \Tilde{\Sigma}}[-(r^{2}+a^{2})A_{1}(r)E+aA_{2}(r)J_{z}]^{2}+\frac{1}{\Tilde{\Sigma} \sin^{2}\theta}[A_{3}(\theta)J_{z}-aA_{4}(\theta)E\sin^{2}\theta]^{2}\\ & +\frac{\Delta}{\Tilde{\Sigma}}A_{5}(r)\Big( \frac{\partial S_{r}}{\partial r}\Big)^{2}+\frac{1}{\Tilde{\Sigma}}A_{6}(\theta)\Big( \frac{\partial S_{\theta}}{\partial \theta}\Big)^{2}\,.
\end{split}
\end{align}
The point of writing the above equation is to show that even after introducing some $r$ and $\theta$ dependent deviation functions, one can still separate the equations and show the existence of a Carter-like constant. $\tilde{\Sigma}$ is a function of $r$ and $\theta$ both. One can easily read off contravariant and covariant forms of the metric and impose further conditions to constrain the form of the above deviation functions in the following way \cite{Johannsen:2013szh}: regularity of the metric gives $A_{5}(r)>0$, the conditions for asymptotic flatness gives $A_{6}(\theta)=1$. Also, one can expand the deviation functions in powers of $\frac{M}{r}$ with coefficients of each term given by $\alpha_{i,n}$ where $i=1,\cdots n$ denoting the power of $\frac{M}{r}$ and $n=1,\cdots 5$ denote the deviation functions. If one expands the metric in $\frac{1}{r}$, $A_{3}(\theta), A_{4}(\theta)$ get constrained to 1 \cite{Johannsen:2013szh}. The deviation parameters can also be further constrained using PPN constraints\footnote{The reader is encouraged to look into \cite{Johannsen:2013szh} for further details.}. Summarizing all of these and keeping only the next-to-leading order term in the metric expansion, the metrics take the following form \cite{Johannsen:2013szh, Staelens:2023jgr}:
\begin{align}\label{metric}
ds^{2} =& -\tilde{\Sigma}\frac{\Delta-a^{2}A_{2}^{2}\sin^{2}\theta}{N}dt^{2}+\frac{\tilde{\Sigma}}{\Delta A_{5}}dr^{2}+\tilde{\Sigma} d\theta^{2}-2a\tilde{\Sigma}\sin^{2}\theta \frac{(r^{2}+a^{2})A_{1}A_{2}-\Delta}{N}dt d\phi + \\ \nonumber 
& \tilde{\Sigma} \sin^{2}\theta\frac{(r^{2}+a^{2})^{2}A_{1}^{2}-a^{2}\Delta\sin^{2}\theta}{N}d\phi^{2}, 
\end{align}
where $A_{1}, A_{2}, A_{5}$ and $N$ are functions of radial coordinate $r$, given by
\begin{align}
A_{1}(r) =& 1+\alpha_{13}\Big(\frac{M}{r}\Big)^{3} \hspace{7mm} ; \hspace{7mm} N(r) = \Big((r^{2}+a^{2})A_{1}-a^{2}A_{2}\sin^{2}\theta\Big)^{2}\,,\nonumber\\
A_{2}(r) =& 1+\alpha_{22}\Big(\frac{M}{r}\Big)^{2} \hspace{7mm} ; \hspace{7mm} \Delta(r) = r^{2}-2M r+a^{2}\,, \nonumber\\
A_{5}(r) =& 1+\alpha_{52}\Big(\frac{M}{r}\Big)^{2} \hspace{7mm} ; \hspace{7mm} \tilde{\Sigma} = r^{2}+a^{2}\cos^{2}\theta+\epsilon_{3} \frac{M^{3}}{r}, \label{eq2}
\end{align}
where ($\alpha_{13}, \alpha_{22}, \alpha_{52}, \epsilon_{3}$) denote deformation parameters in Kerr. The interesting point about this metric is that the event horizon and the multipole moments coincide with that of the Kerr. 

 Note that, following  \cite{Johannsen_2010, Johannsen1_2010, Johannsen_2011, PhysRevD.83.124015, Johannsen:2013szh, Carson:2020dez}, we have applied constraints coming from parametrized-post-Newtonian (ppN) framework. Otherwise, we have to deal with more deviation parameters.\footnote{It should be noted that the ppN constraints are obtained assuming the central object is a star. Therefore, to apply these constraints, we must further assume that the asymptotic behaviours of metric components for a star and a black hole geometry are the same \cite{Cardoso:2014rha, Carson:2020dez}.}. Even after using such constraints, the parameters $\alpha_{i,n}s$ as indicated in Eq.(\ref{eq2}) remain unconstrained. One may, in principle, consider large numerical values for these parameters. However, since we perform all our analysis in this paper perturbatively for deviation parameters, assuming that the dominant contribution always comes from GR, we keep them small and consider only corresponding leading-order PN contributions.

\section{Geodesic motion and eccentric orbital dynamics}\label{gdscvecmtn}
In this section we discuss the leading order effects of Johannsen parameters (deviation parameters to the Kerr) as defined in Eq.(\ref{eq2}) on the eccentric equatorial orbital motion of the inspiralling object. Since, the metric has two Killing vectors ($\partial_{t})^{\mu}$ and ($\partial_{\phi})^{\mu}$, it gives rise to two conserved quantities energy ($E$) and angular momentum ($J_{z}$). We also have the third constant of motion, the Carter constant. Since we focus on equatorial motion ($\theta=\pi/2$), the  Carter constant will be zero ($\mathcal{Q}=0$) \cite{Glampedakis:2002cb, Glampedakis:2002ya, PhysRevD.61.084004}. It is always good to introduce dimensionless quantities: $\hat{r}=r/M, \hat{a}=a/M, \hat{J}_{z}=J_{z}/(\mu M)$ and $\hat{E}=E/\mu$; 
We will further set $M=1$. For writing convenience, we avoid putting the hat over the aforesaid quantities. With this, the equations of motion of the object upto $\mathcal{O}(a)$ can be written in the following way

\begin{equation}
\begin{aligned}\label{gdscs2n}
\frac{dt}{d\tau} =& \Big[-(r^{2}A_{1}A_{2}-\Delta)\frac{aJ_{z}}{r^{2}\Delta}+\frac{Er^{2}}{\Delta}A_{1}^{2}\Big]\Big(1+\frac{\epsilon_{3}}{r^{3}}\Big)^{-1} + \mathcal{O}(a^{2})\\
\frac{d\phi}{d\tau} =& \Big[\frac{J_{z}}{r^{2}\sin^{2}\theta}+\frac{aE}{r^{2}\Delta}(A_{1}A_{2}r^{2}-\Delta)\Big]\Big(1+\frac{\epsilon_{3}}{r^{3}}\Big)^{-1}+ \mathcal{O}(a^{2}) \\
\Big(\frac{dr}{d\tau}\Big)^{2} =& \frac{A_{5}}{\Sigma^{2}}\Big[(A_{1}r^{2}E-aA_{2}J_{z})^{2}-\Delta\Big(Q+r^{2}+\frac{\epsilon_{3}}{r}\Big)\Big] + \mathcal{O}(a^{2}) \\
\Big(\frac{d\theta}{d\tau}\Big)^{2} =& \frac{1}{\Sigma^{2}}\Big[Q-\Big(aE\sin\theta-\frac{J_{z}}{\sin\theta}\Big)^{2}\Big] + \mathcal{O}(a^{2}),
\end{aligned}
\end{equation}
where, $\mathcal{Q}\equiv Q-(J_{z}-aE)^{2}$. Since $\mathcal{Q}=0$, it implies that $Q=J_{z}^{2}-2aEJ_{z}+\mathcal{O}(a^{2})$ as also discussed in appendix (\ref{apenteu1}). We further analyze the radial equation in Eq.(\ref{gdscs2n}), which governs the object's orbital motion and provides us with the effective potential ($V_{eff}$). In general, without loss of generality, one can also express the effective potential for equatorial orbits in terms of metric coefficients as \cite{Rahman:2023sof}
\begin{align}\label{vef1}
\Big(\frac{dr}{d\tau}\Big)^{2} = - V_{eff}(r), \hspace{3mm}\textup{where}, \hspace{3mm} \hspace{3mm} V_{eff}(r) = \frac{1}{2}(g^{tt}E^{2}-2g^{t\phi}EJ_{z}+g^{\phi\phi}J_{z}^{2}+1).
\end{align}
Note that we keep only upto leading order terms in spin and Johannsen parameters ($a, \alpha_{13}, \alpha_{52}, \alpha_{22}, \epsilon_{3}$) and discard subleading corrections. 

As we aim to examine the eccentric dynamics of the object, we take two turning points ($r_{a}, r_{p}$) for bound orbits where the radial velocity vanish ($V_{eff}=0$), denoting \textit{apastron} and \textit{periastron} respectively. The bound orbits can be written in terms of eccentricity ($e$) and semi-latus rectum ($p$) as
\begin{align}\label{bd1}
r_{a} = \frac{p}{1-e} \hspace{5mm} ; \hspace{5mm} r_{p} = \frac{p}{1+e}\,.
\end{align}
The condition for bound orbits in the range $r_{p}<r<r_{a}$ is $V_{eff}(r)<0$, which is satisfied only when $V'_{eff}(r_{a})>0$ and $V'_{eff}(r_{p})\leq 0$ \cite{Skoupy:2021asz}; where, prime denotes the derivative with respect to $r$. Using Eq.(\ref{vef1}) and Eq.(\ref{bd1}), at the turning points for the system under consideration, one can derive the constants of motion of the bound orbits; the positive roots of ($E, J_{z}$) read as
\begin{align}\label{cnst1}
E =& \sqrt{\frac{(p-2)^2-4 e^2}{p \left(p-3-e^2\right)}} +\frac{(e^2-1)^2}{p\left(p-3-e^2\right)^{3/2}}\Big[-a +\frac{\alpha_{13} \sqrt{(p-2)^2-4 e^2}}{2 p^{3/2}}\Big]-\frac{\epsilon_{3}(e^{2}-1)^{2}}{4} \sqrt{\frac{(p-2)^2-4 e^2}{p^{7}(p-3-e^2)}}\,, \\
J_{z} =&\frac{p}{\sqrt{p-3-e^2}}+(3+e^{2})\sqrt{\frac{(p-2)^{2}-4e^{2}}{p(p-3-e^{2})^{3}}}\Big[-a+\frac{\alpha_{13}\sqrt{(p-2)^{2}-4e^{2}}}{2p^{3/2}}\Big] -\frac{\epsilon_{3} \left(e^2 (p-8)+3 p-8\right)}{4 p^2 \sqrt{p-3-e^2}}. \label{cnst2}
\end{align}
We remind that we only consider leading order corrections in spin and Johannsen parameters through out the article. Once we have the constants of motion, one can always determine the last stable orbit (LSO)  of the inspiralling object \cite{PhysRevD.103.104045, Glampedakis:2002cb}. This requires the following condition to satisfy: $V_{eff}(r_{p})=0, V_{eff}(r_{a})=0$; $V'_{eff}(r_{p})=0, V'_{eff}(r_{a})>0$. From Eqs.(\ref{vef1}, \ref{bd1}, \ref{cnst1}, \ref{cnst2}), one can obtain the set of ($p, e$) points that will separate the bound orbits from unbounded ones \cite{PhysRevD.50.3816, PhysRevD.77.124050}. We term such a region as \textit{separatrix}, and the expression takes the following form
\begin{align}\label{sp1}
p_{sp}=2 (e+3)-4 \sqrt{2} a \sqrt{\frac{e+1}{e+3}}+4 \alpha_{13} \frac{(e+1)}{(e+3)^2}-\epsilon_{3}\frac{(e-3)^2 (e+1)^2}{4 (e+3)^3}\,.
\end{align}
To put it another way, the separatrix determines the lowest semi-latus rectum value for which spacetime permits bound orbits for a given eccentricity. It is the last stable orbit of the motion. For circular orbits ($e=0$), it is called the innermost stable circular orbit (ISCO). The Eq.(\ref{sp1}) corresponds to the prograde motion of the inspiralling object and coincides with \cite{Glampedakis:2002ya} in leading order $a$; however, one can replace $a\rightarrow -a$ for the retrograde motion. It implies that the separatrix curve for prograde (retrograde) orbits will shift to the left (right) with respect to the Schwarzschild curve ($p^{sch}_{sp}=6+2e$) as the black hole spins up \cite{Glampedakis:2002ya}. 

The motion occurs between $r_{p}$ to $r_{a}$ and vice-versa. As we choose $r$ as the parameter throughout the orbit, we can integrate Eq.(\ref{gdscs2n}) by removing $\tau$ from the set of equations. We parametrize the radial coordinate $r$ in the following way to conquer the divergences at the turning points ($V_{eff}=0; r_{a}, r_{p}$)
\begin{align}\label{prmtz}
r=\frac{p}{1+e\cos\chi}\,.
\end{align}
This parametrization helps in removing the singular behaviour at the turning points ($r_{a}, r_{p}$) in the differential equations, i.e., in geodesic velocities, implying ($\chi=\pi, \chi=0$) respectively. Further, the eccentric motion exhibits two fundamental frequencies: radial ($\Omega_{\phi}$) and azimuthal ($\Omega_{r}$). It has been shown that the radial motion shows the periodicity, not the azimuthal one \cite{PhysRevD.50.3816}; hence, using Eq.(\ref{gdscs2n}), ($\Omega_{\phi}, \Omega_{r}$) are given by
\begin{align}
\Omega_{\phi} =& \frac{1}{E r^3}\Big(J_{z} (r-2)+\frac{a \left(2 E^2 r^3+2 J_{z}^2 (r-2)\right)}{E r^3}-\frac{2 \alpha_{13} J_{z} (r-2)}{r^3}\Big), \nonumber \\
\Omega_{r} =& \frac{2\pi}{T_{r}} \hspace{3mm} ; \hspace{3mm} T_{r} = \int_{0}^{2\pi}d\chi\frac{dt}{d\chi} \hspace{3mm} ; \hspace{3mm} \frac{dt}{d\chi} = \frac{dt}{dr}\frac{dr}{d\chi}.
\end{align}
In general, $\Omega_{\phi}$ can be written in terms of metric coefficients: $\Omega_{\phi}=-\frac{g_{t\phi}E+g_{tt}J_{z}}{g_{\phi\phi}E+g_{t\phi}J_{z}}$, that coincides in linear order corrections with \cite{Carson:2020dez}. $T_{r}$ is the radial time period. We use these relevant quantities later in subsequent sections for estimating fluxes. 


\section{Radiation reaction: GW fluxes and orbital evolution} \label{radiation reaction}

Let us turn our discussion to the GW radiation reaction acceleration or force that gradually alters the orbital dynamics of the inspiralling object. As we consider an EMRI system, the primary supermassive black hole is being perturbed by the secondary object that plays the role of a test particle. As a result, emitted GWs introduce the notion of GW radiation reaction, which infers that the constants of motion no longer remain constants; they start evolving in time. Although the overall scenario necessitates a complete numerical analysis \cite{Hughes:1999bq}, we analyze effects within the leading order PN corrections as well as leading order in the mass ratio. We mainly follow the analysis of \cite{Flanagan:2007tv, PhysRevD.52.R3159, Ryan:1995xi}. In this section, we provide a general setup to estimate the average energy and angular momentum fluxes due to the radiation reaction effects for eccentric equatorial orbits and examine the leading order deviations from the Kerr results.  

It is useful to introduce Cartesian coordinates ($x_{1},x_{2},x_{3})=(r\sin\theta \cos\phi, r\sin\theta \sin\phi, r\cos\theta$) for expressing the constants of motion and their change. The conserved quantities in the leading order of deviation parameters have been derived in Eqs.(\ref{ener2}, \ref{dphidt}, \ref{cart1}) and are given in Cartesian coordinates as
\begin{align}
\mathcal{E} =& \frac{1}{2} \dot{x}_{i}\dot{x}_{i}-\frac{1}{\sqrt{x_{i}x_{i}}}, \\
J_{z} =& \epsilon_{3jk}x_{j}\dot{x}_{k}\Big(1+\frac{\epsilon_{3}}{(x_{i}x_{i})^{3/2}}\Big)-2a\frac{(x_{1}^{2}+x_{2}^{2})}{(x_{i}x_{i})^{3/2}}, \\
\mathcal{Q}+J_{z}^{2} =& (\epsilon_{ijk}x_{j}\dot{x}_{k})(\epsilon_{ilm}x_{l}\dot{x}_{m})-4a\frac{\epsilon_{3jk}x_{j}\dot{x}_{k}}{\sqrt{x_{i}x_{i}}},
\end{align}
where $r^{2}\sin^{2}\theta \dot{\phi}=\epsilon_{3jk}x_{j}\dot{x}_{k}$ and $r^{4}(\dot{\theta}^{2}+\sin^{2}\theta\dot{\phi}^{2})=(\epsilon_{ijk}x_{j}\dot{x}_{k})(\epsilon_{ilm}x_{l}\dot{x}_{m})$, and dot denotes the derivative with respect to coordinate time $t$. Note that the analysis explained here is also valid for non-equatorial orbits and will match with the results of \cite{Flanagan:2007tv} up to linear order in $a\,.$ We will set $\theta=\pi/2$ at a later stage while computing the average fluxes for the ease of the computation. The radiation reaction effect causes the perturbed system to generate instantaneous fluxes, i.e., the rate change of constants of motion takes the following form
\begin{align}\label{inst fluxes}
\dot{\mathcal{E}} = x_{i}\Ddot{x}_{i} \hspace{2mm} ; \hspace{2mm}
\dot{J}_{z} = \epsilon_{3jk}x_{j}\Ddot{x}_{k}\Big(1+\frac{\epsilon_{3}}{(x_{i}x_{i})^{3/2}}\Big) \hspace{2mm} ; \hspace{2mm}
\dot{\mathcal{Q}}+\dot{(J_{z}^{2})} = 2(\epsilon_{ijk}x_{j}\dot{x}_{k})(\epsilon_{ilm}x_{l}\ddot{x}_{m})-\frac{4a\epsilon_{3jk}x_{j}\Ddot{x}_{k}}{\sqrt{x_{i}x_{i}}}.
\end{align}
Since we are interested in analyzing the contributions of the radiative part, we only consider terms involving acceleration, also called radiation reaction acceleration and often denoted by $a_{j}$. The general expression of the same is given as \cite{Flanagan:2007tv, PhysRevD.52.R3159}
\begin{align}\label{accelration}
a_j=-\dfrac{2}{5}I^{(5)}_{jk}x_{k}+\dfrac{16}{45}\epsilon_{jpq}J^{(6)}_{pk}x_{q}x_{k}+\dfrac{32}{45}\epsilon_{jpq}J^{(5)}_{pk}x_{k} \dot{x}_{q}+\dfrac{32}{45}\epsilon_{pq[j}J^{(5)}_{k]p}x_{q} \dot{x}_{k}+\dfrac{8J}{15}J^{(5)}_{3i},
\end{align}
where $I_{jk}$ and $J_{jk}$ are termed as mass and current quadrupole moments. The last term with $J$ is the spin parameter ($a$) of the central black hole. The superscripts define the order of derivative of a quantity. For the third term, the anti-symmetric quantity can written as $B_{[ij]}=\frac{1}{2}(B_{ij}-B_{ji})$. 
We take the symmetric trace-free (STF) part of the mass and current moments in the following manner
\begin{equation}\label{moments}
I_{jk}=\Big[x_j x_k\Big]^{\text{STF}} \hspace{3mm} ; \hspace{3mm}
J_{jk}=\Big[x_{j}\epsilon_{kpq}x_{p}\dot{x}_{q}-\dfrac{3}{2}x_jJ\delta_{k3}\Big]^{\text{STF}}.
\end{equation}
We compute the quantities mentioned in Eq.(\ref{moments}) and replace them in the Eq.(\ref{accelration}). Further, to estimate the instantaneous change of the constants of motion, we make use of Eq.(\ref{accelration}) and Eq.(\ref{inst fluxes}). Since we focus on the equatorial orbital motion, the expression in Eq.(\ref{inst fluxes}) involving the Carter constant will not be considered for computing the evolution of constants of motion as $\theta=\pi/2$. 

The radiation reaction force causes the constants of motion to evolve in time. We use velocities (\ref{gdscs3n}) and calculate these quantities, and further average them out. We perform that computation upto the leading order  in ($a, \alpha_{13}, \epsilon_{3}, \alpha_{52}, \alpha_{22}$). This also sets up the platform to analyze the effects of Johannsen parameters in the orbital evolution of the inspiralling object. The instantaneous fluxes or rate change of constants of motion are given as

\begin{equation}
\begin{aligned} \label{instflx}
\dot{\mathcal{E}} =& \frac{272 J_{z}^2 \mathcal{E}}{5 r^5}+\frac{160 J_{z}^2}{3 r^6}-\frac{40 J_{z}^4}{r^7}+a \left(\frac{196 J_{z}^5}{r^9}-\frac{3668 J_{z}^3}{5 r^8}-\frac{352 J_{z}^3 \mathcal{E}}{r^7}+\frac{952 J_{z}}{3 r^7}+\frac{1024 J_{z} \mathcal{E}}{3 r^6}+\frac{128 J_{z} \mathcal{E}^2}{5 r^5}\right)\\
&+\alpha_{52} \left(\frac{2048 J_{z}^6}{5 r^{10}}-\frac{20192 J_{z}^4}{15 r^9}-\frac{4352 J_{z}^4 \mathcal{E}}{5 r^8}+\frac{5696 J_{z}^2}{5 r^8}+\frac{19504 J_{z}^2 \mathcal{E}}{15 r^7}+\frac{256 J_{z}^2 \mathcal{E}^2}{r^6}-\frac{320}{3 r^7}-\frac{1216 \mathcal{E}}{5 r^6}-\frac{128 \mathcal{E}^2}{r^5}\right)\\
&+\alpha_{13} \left(-\frac{392 J_{z}^4}{r^9}+\frac{4256 J_{z}^2}{5 r^8}+\frac{752 J_{z}^2 \mathcal{E}}{r^7}-\frac{320}{3 r^7}-\frac{1216 \mathcal{E}}{5 r^6}-\frac{128 \mathcal{E}^2}{r^5}\right)+\epsilon_{3} \Big(-\frac{126 J_{z}^6}{r^{11}}+\frac{342 J_{z}^4}{5 r^{10}}+\frac{196 J_{z}^4}{r^9}\\
&+\frac{7412 J_{z}^2}{15 r^9}+\frac{4344 J_{z}^2 \mathcal{E}}{5 r^8}-\frac{2128 J_{z}^2}{5 r^8}+\frac{384 J_{z}^2 \mathcal{E}^2}{r^7}-\frac{376 J_{z}^2 \mathcal{E}}{r^7} +\frac{160}{3 r^7}+\frac{608 \mathcal{E}}{5 r^6}+\frac{64 \mathcal{E}^2}{r^5}+\frac{64}{3 r^5}+\frac{512 \mathcal{E}}{15 r^4}+\frac{64 \mathcal{E}^2}{5 r^3}\Big)\,, \\ \\
\dot{J}_{z} =&\frac{144 J_{z} \mathcal{E}}{5 r^3}+\frac{32 J_{z}}{r^4}-\frac{24 J_{z}^3}{r^5}+ a \left(\frac{140 J_{z}^4}{r^7}-\frac{7004 J_{z}^2}{15 r^6}+\frac{8 \left(395-474 J_{z}^2 \mathcal{E}\right)}{15 r^5}+\frac{1248 \mathcal{E}}{5 r^4}+\frac{256 \mathcal{E}^2}{5 r^3}\right) \\
&+\alpha_{13} \left(-\frac{216 J_{z}^3}{r^7}+\frac{2016 J_{z}}{5 r^6}+\frac{336 J_{z} \mathcal{E}}{r^5}\right)+\alpha_{52} \Big(\frac{1056 J_{z}^5}{5 r^8}-\frac{608 J_{z}^3}{r^7}+\frac{336 J_{z} \mathcal{E}}{r^5} \\
& +\frac{288 \left(7 J_{z}-6 J_{z}^3 \mathcal{E}\right)}{5 r^6}\Big) +\epsilon_{3} \Big(-\frac{42 J_{z}^5}{r^9}-\frac{666 J_{z}^3}{5 r^8}+\frac{12 \left(-10 J_{z}^3 \mathcal{E}+9 J_{z}^3+37 J_{z}\right)}{r^7} \\
& +\frac{24 (151 J_{z} \mathcal{E}-42 J_{z})}{5 r^6} +\frac{24 \left(12 J_{z} \mathcal{E}^2-7 J_{z} \mathcal{E}\right)}{r^5}\Big).
\end{aligned}
\end{equation}
In principle, the motion of the stable orbits is restricted to a region whose shape can be determined by ($\mathcal{E}, J_{z}, \mathcal{Q}+J_{z}^{2}$). An equivalent way to describe the motion is to specify ($p, e, \iota$) \cite{PhysRevD.61.084004}, where $\iota$ is the inclination angle. Also we have: ${r_{p,a}=\frac{p}{1\pm e}}$ and $\cos\iota = \frac{J_{z}}{(\mathcal{Q}+J_{z}^{2})^{1/2}}$; $r_{p,a}$ are the turning points. Since we are considering equatorial orbits, the motion will be characterized by ($\mathcal{E}, J_{z}$) or equivalently ($p, a$) as $\iota=0$ and $\mathcal{Q}=0$. Then using radial equation Eq.(\ref{gdscs3n}), at the turning points ($\mathcal{E}, J_{z}$) become
\begin{equation}
\begin{aligned}\label{enrjz2}
\mathcal{E} =& \frac{(1-e^{2})}{p}\Big[-\frac{1}{2}-\frac{a \left(1-e^2\right)}{p^{3/2}}+\frac{\alpha_{13} \left(1-e^2\right)}{2 p^2}-\frac{ \epsilon_{3}\left(1-e^2\right)}{4p^{2}}  \Big]\,, \\
J_{z} =& \sqrt{p}-\frac{a \left(e^2+3\right)}{p}+\frac{\alpha_{13} \left(e^2+3\right)}{2 p^{3/2}}+\frac{\alpha_{52} \left(1-e^2\right)}{2 p^{3/2}}-\frac{\epsilon_{3}(e^{2}+3)}{4p^{3/2}}.
\end{aligned}
\end{equation}
These results, in linear order $a$, coincide with the ones presented in \cite{Flanagan:2007tv} without the deviation parameters. Here, we take leading order contributions from the deviation parameters and discard subleading terms. Also, we make use of $E=\mu+\mathcal{E}$ as discussed in Appendix (\ref{apenteu1}). Plugging Eq.(\ref{enrjz2}) into Eq.(\ref{instflx}), the instantaneous fluxes become

\begin{equation}
\begin{aligned}\label{instflx2}
\dot{\mathcal{E}}=& \frac{16 e^4}{5 p^2 r^3}-\frac{32 e^2}{5 p^2 r^3}+\frac{256 e^2}{15 p r^4}+\frac{136 e^2}{5 r^5}+\frac{16}{5 p^2 r^3}-\frac{256}{15 p r^4}-\frac{88}{15 r^5}+\frac{160 p}{3 r^6}-\frac{40 p^2}{r^7} \\
& +a \Big(-\frac{3668 p^{3/2}}{5 r^8}+\frac{196 p^{5/2}}{r^9}+\frac{2920 \sqrt{p}}{3 r^7}-\frac{64 e^6}{5 p^{7/2} r^3}+\frac{192 e^4}{5 p^{7/2} r^3}-\frac{192 e^2}{5 p^{7/2} r^3}+\frac{64}{5 p^{7/2} r^3} \\
& -\frac{512 e^4}{15 p^{5/2} r^4}+\frac{1024 e^2}{15 p^{5/2} r^4}-\frac{512}{15 p^{5/2} r^4}-\frac{512 e^4}{5 p^{3/2} r^5}-\frac{64 e^2}{5 p^{3/2} r^5}+\frac{576}{5 p^{3/2} r^5} +\frac{64 e^2}{\sqrt{p} r^6} -\frac{1472}{3 \sqrt{p} r^6} \\
& -\frac{16 e^2 \sqrt{p}}{r^7}\Big)+\alpha_{13}\Big(\frac{32 e^6}{5 p^4 r^3}-\frac{96 e^4}{5 p^4 r^3}+\frac{256 e^4}{15 p^3 r^4}+\frac{112 e^4}{5 p^2 r^5}+\frac{96 e^2}{5 p^4 r^3}-\frac{512 e^2}{15 p^3 r^4} +\frac{64 e^2}{p^2 r^5} -\frac{1024 e^2}{15 p r^6} \\
& +\frac{296 e^2}{r^7}-\frac{32}{5 p^4 r^3}+\frac{256}{15 p^3 r^4}-\frac{432}{5 p^2 r^5}+\frac{1408}{5 p r^6}-\frac{2168}{3 r^7}+\frac{4256 p}{5 r^8}-\frac{392 p^2}{r^9}\Big) +\alpha_{52}\Big(-\frac{296 e^4}{5 p^2 r^5} \\
& +\frac{64 e^4}{p r^6} +\frac{592 e^2}{5 p^2 r^5}-\frac{4544 e^2}{15 p r^6}+\frac{10952 e^2}{15 r^7}-\frac{2176 p e^2}{5 r^8}-\frac{296}{5 p^2 r^5}+\frac{3584}{15 p r^6}-\frac{4184}{5 r^7} +\frac{7872 p}{5 r^8} \\
& -\frac{20192 p^2}{15 r^9}  +\frac{2048 p^3}{5 r^{10}}\Big) +\epsilon_{3}\Big(\frac{32 e^8}{5 p^5 r^3}-\frac{16 e^6}{5 p^4 r^3}+\frac{256 e^6}{15 p^4 r^4}+\frac{272 e^6}{5 p^3 r^5}+\frac{48 e^4}{5 p^4 r^3}-\frac{192 e^4}{5 p^5 r^3} -\frac{128 e^4}{15 p^3 r^4} \\
& +\frac{256 e^4}{15 p^4 r^4}-\frac{56 e^4}{5 p^2 r^5} +\frac{272 e^4}{p^3 r^5} +\frac{160 e^4}{3 p^2 r^6}+\frac{16 e^4}{p r^7}-\frac{48 e^2}{5 p^4 r^3}+\frac{256 e^2}{5 p^5 r^3}+\frac{256 e^2}{15 p^3 r^4}-\frac{256 e^2}{3 p^4 r^4} -\frac{32 e^2}{p^2 r^5} \\
& -\frac{272 e^2}{p^3 r^5}+\frac{512 e^2}{15 p r^6}+\frac{1600 e^2}{3 p^2 r^6}-\frac{148 e^2}{r^7} -\frac{992 e^2}{p r^7}+\frac{2172 e^2}{5 r^8}+\frac{16}{5 p^4 r^3}-\frac{96}{5 p^5 r^3} -\frac{128}{15 p^3 r^4} +\frac{256}{5 p^4 r^4} \\
& +\frac{216}{5 p^2 r^5}-\frac{272}{5 p^3 r^5}-\frac{704}{5 p r^6}+\frac{800}{3 p^2 r^6}+\frac{1084}{3 r^7} -\frac{304}{p r^7}-\frac{2128 p}{5 r^8}-\frac{2172}{5 r^8} +\frac{196 p^2}{r^9} +\frac{7412 p}{15 r^9} \\
& +\frac{342 p^2}{5 r^{10}}-\frac{126 p^3}{r^{11}}\Big)\,,
\end{aligned}
\end{equation}
\begin{equation}
\begin{aligned}\label{instflx3}
\dot{J}_{z} =& \frac{24}{5 p^{7/2} r^9}\Big[3 \left(e^2-1\right) p^3 r^6-5 p^5 r^4+20 p^4 r^5\Big]+\frac{a}{15 p^{7/2} r^9}\Big[-24 \left(e^2-1\right) \left(19 e^2+17\right) p^{3/2} r^6 \\
& +48 \left(29 e^2-69\right) p^{5/2} r^5-136 \left(6 e^2-61\right) p^{7/2} r^4 -7004 p^{9/2} r^3+2100 p^{11/2} r^2\Big] \\
& +\frac{\alpha_{13}}{15 p^{7/2} r^9}\Big[(90 \left(22 e^2-46\right) p^3 r^4+240 \left(e^2+3\right) p^2 r^5+54 \left(e^2-1\right) \left(6 e^2+2\right) p r^6 -3240 p^5 r^2 \\
& +6048 p^4 r^3)\Big]+\frac{4\alpha_{52}}{15 p^{7/2} r^9}\Big[108 \left(20-6 e^2\right) p^4 r^3+765 \left(e^2-1\right) p^3 r^4-60 \left(e^2-1\right) p^2 r^5 \\
& -27 p r^6 (1-e^{2})^{2}+792 p^6 r-2280 p^5 r^2\Big]+\frac{\epsilon_{3}}{15 p^{7/2} r^9}\Big[180 \left(42-5 e^2\right) p^4 r^2 +90 \left(23-11 e^2\right) p^3 r^4 \\
& +5436 \left(e^2-1\right) p^3 r^3-120 \left(e^2+3\right) p^2 r^5-162 e^2 \left(e^2-1\right) p r^6-54 \left(e^2-1\right) p r^6 \\
& +540 \left(e^4-14 e^2-3\right) p^2 r^4+240 \left(e^4+10 e^2+5\right) p r^5 +108 \left(e^2-1\right) \left(3 e^4+14 e^2-1\right) r^6 \\
& -630 p^6+1620 p^5 r^2-1998 p^5 r-3024 p^4 r^3\Big].
\end{aligned}
\end{equation}
With the instantaneous fluxes in hand, let us now determine the average rate change of these quantities. We take the average within adiabatic approximation \cite{Hinderer:2008dm, PhysRevD.103.104014, PhysRevLett.128.231101}, where it is justifiable to consider the motion of the inspiralling object on a geodesic. In other words, as long as we are concerned with timescales much shorter than the radiation reaction time scale, the adiabatic approximation is useful to consider the motion of the particle to be geodesic \cite{Glampedakis:2002ya}. The radiative losses will cause ($\dot{\mathcal{E}}, \dot{J_{z}}$) to vary slowly, and ($<\dot{\mathcal{E}}>, <\dot{J}_{z}>$) are estimated by taking the average over one orbit. Thus, in general, the expression for average fluxes can be written as
\begin{align}\label{flx1}
<\dot{C}> = \frac{1}{T_{r}}\int^{2\pi}_{0}\dot{C}(\chi)\frac{dt}{d\chi} d\chi \hspace{5mm} ; \hspace{5mm} C \equiv (\mathcal{E}, J_{z}).
\end{align}
 It is to be noted that higher-order PN terms (at 3PN order) proportional to $\epsilon_{3}$ coefficient coming from Eq.(\ref{instflx2}, \ref{instflx3}) after averaging are discarded (these are associated with a $\frac{1}{p^{8}}$ factor in energy flux and $\frac{1}{p^{13/2}}$ in the angular momentum flux.) as we are at this point only interested in the leading order contribution coming from the deviation parameters. As we will discuss below, the leading order contribution will appear at the 2PN order. We similarly discard this higher-order correction in orbital evolution, phase and dephase. For calculating $T_{r}$, we make use of parametrization Eq.(\ref{prmtz}) and radial velocity; the time period with the leading corrections reads as
\begin{align}\label{tmprd}
T_{r} = \frac{\pi}{\sqrt{1-e^{2}}}\Big[\frac{2  p^{3/2}}{\left(1-e^2\right)}-6a+\frac{1}{\sqrt{p}}\Big(3\alpha_{13}-\alpha_{52}
-\frac{3\epsilon_{3}}{2}\Big)\Big].
\end{align}
Using Eqs.(\ref{instflx2}, \ref{instflx3}) and Eq.(\ref{tmprd}), the average loss of energy and angular momentum fluxes are given by
\begin{equation}
\begin{aligned}\label{avgflx1}
\Big\langle\frac{d\mathcal{E}}{dt}\Big\rangle =& -\frac{\left(1-e^2\right)^{3/2}}{15 p^{5}}\Big[\left(37 e^4+292 e^2+96\right)-\Big\{\frac{a}{p^{3/2}} \left(\frac{491 e^6}{2}+2847 e^4+3292 e^2+584\right) \\
& -\frac{\alpha_{13}}{p^2}\left(251 e^6+4035 e^4+5532 e^2+864\right) +\frac{\alpha_{52}}{p^2} \Big(\frac{2775 e^6}{4}+2275 e^4-130 e^2-98e^{2} \sqrt{1-e^2} \\
& -48 \sqrt{1-e^2}+\frac{37e^{6} \sqrt{1-e^2}}{2}+\frac{255e^{4} \sqrt{1-e^2}}{2}-240\Big)+\frac{\epsilon_{3}}{p^{2}} \Big(\frac{251 e^6}{2} +\frac{4035 e^4}{2}+2766 e^2 + 432\Big)\Big\}\Big],
\end{aligned}
\end{equation}
\begin{equation}
\begin{aligned}\label{avgflx2}
\Big\langle\frac{dJ_{z}}{dt}\Big\rangle =& -\frac{(1-e^{2})^{3/2}}{5p^{3}}\Big[\frac{4}{p^{1/2}}\left(7 e^2+8\right)-\frac{a}{3 p^{2}} \left(549 e^4+1428 e^2+488\right)+\frac{2\alpha_{13}}{p^{5/2}} \left(84 e^4+361 e^2+120\right) \\
& -\frac{\alpha_{52}}{2(1-e^{2})^{1/2}p^{5/2}} \Big\{\left(-28 e^6+24 e^4+36 e^2-32\right)+\sqrt{1-e^2} \left(12 e^6+421 e^4+192 e^2-128\right)\Big\} \\
& -\frac{\epsilon_{3}}{p^{5/2}} \Big(84 e^4 + 361e^2 + 120\Big)\Big].
\end{aligned}
\end{equation}
Expressions Eq.(\ref{avgflx1}) and Eq.(\ref{avgflx2}) show average energy and angular momentum fluxes in the leading order corrections of spin and deviation parameters, which match with the standard results of \cite{Flanagan:2007tv, Glampedakis:2002ya, Ryan:1995xi, PhysRev.131.435, PhysRev.136.B1224} in the equatorial limit when the deviation parameters are turned off. We know that the leading order term coming from GR, in various fluxes, appears with a $\mathcal{O}(c^{-5})$ factor \cite{PhysRev.131.435}\footnote{Restoring the factors of $c$, we can see that the expressions for the fluxes take the following form: $$\langle\frac{d\mathcal{E}}{dt}\rangle  \sim -\frac{1}{p^5\,c^{5}}(a_1+a_2\frac{a}{p^{3/2}\,c^{3}}+a_3\frac{\alpha_{13}}{p^2\,c^{4}}+a_4\frac{\alpha_{52}}{p^2\,c^{4}}+a_5\frac{\epsilon_{3}}{p^2\,c^{4}}),\, \langle\frac{dJ_{z}}{dt}\rangle\sim -\frac{1}{p^3\,c^{5}}(a_6\frac{1}{p^{1/2}}+a_7\frac{a}{p^{2}\,c^{3}}+a_8\frac{\alpha_{13}}{p^{5/2}\,c^{4}}+a_9\frac{\alpha_{52}}{p^{5/2}\,c^{4}}+a_{10}\frac{\epsilon_{3}}{p^{5/2}\,c^{4}})\,.$$ After pulling out an overall factor of $\frac{1}{c^5}$ we can identify the  PN order $n\,,$ $(\frac{v}{c})^{2n}$ at which the deviation parameters contribute to the fluxes.$a_i, i=1,\cdots 10$ are just some eccentricity ($e$) dependant constant pre-factors.}. Then, one can easily check that, after restoring the factors of $c,$ the leading correction due to the spin from the GR part shows up at 1.5PN order, and the correction due to the deviation parameters first appears at the 2PN order. Let us now examine the orbital evolution of the inspiralling object. Since our analysis holds within the adiabatic approximation, one can write down the following balance equations \cite{Hughes:2021exa}
\begin{align}\label{blnceqn}
\Big\langle\frac{d\mathcal{E}}{dt}\Big\rangle_{GW} = -\Big\langle\frac{d\mathcal{E}}{dt}\Big\rangle \hspace{5mm} ; \hspace{5mm} \Big\langle\frac{dJ_{z}}{dt}\Big\rangle_{GW} = -\Big\langle\frac{dJ_{z}}{dt}\Big\rangle.
\end{align}
This means that, in the adiabatic approximation, the averaged radiated energy and angular momentum flux are equivalent to the orbital energy and angular momentum loss. It provides the GW fluxes that match with \cite{Dai:2023cft, PhysRev.131.435}. We use Eq.(\ref{avgflx1}) and Eq.(\ref{avgflx2}) for computing the average rate change of eccentric orbital parameters, which can be obtained from the following relation
\begin{align}\label{dpdtdedt1}
\frac{d\mathcal{E}}{dt} = \frac{\partial \mathcal{E}}{\partial p}\frac{dp}{dt} + \frac{\partial\mathcal{E}}{\partial e}\frac{de}{dt} \hspace{5mm} ; \hspace{5mm} \frac{dJ_{z}}{dt} = \frac{\partial J_{z}}{\partial p}\frac{dp}{dt} + \frac{\partial J_{z}}{\partial e}\frac{de}{dt}.
\end{align}
Here, we use the average loss of energy and angular momentum fluxes. Then we get,
\begin{align}\label{dpdtdedt2}
\Big\langle\frac{dp}{dt}\Big\rangle = \Big(\frac{\dot{\mathcal{E}}\partial_{e}J_{z} -\dot{J}_{z}\partial_{e}\mathcal{E}}{\partial_{p}\mathcal{E}\partial_{e}J_{z}-\partial_{e}\mathcal{E}\partial_{p}J_{z}}\Big) \hspace{3mm} ; \hspace{3mm} \Big\langle\frac{de}{dt}\Big\rangle = \Big(\frac{\dot{J}_{z}\partial_{p}\mathcal{E}-\dot{\mathcal{E}}\partial_{p}J_{z}}{\partial_{p}\mathcal{E}\partial_{e}J_{z}-\partial_{e}\mathcal{E}\partial_{p}J_{z}}\Big).
\end{align}
Using Eq.(\ref{instflx}), Eq.(\ref{enrjz2}) and Eq.(\ref{tmprd}) and further taking the average of the resultant expression or directly using Eq.(\ref{avgflx1}) and Eq.(\ref{avgflx2}), the average rate change of ($p, e$) are given by
\begin{equation}
\begin{aligned}\label{dpdtn}
\Big\langle\frac{dp}{dt}\Big\rangle =& -\frac{2(1-e^{2})^{3/2}}{5p^{3}}\Big[4\left(7 e^2+8\right)-\frac{a}{3 p^{3/2}}\left(475 e^4+1516 e^2+1064\right)+\frac{\alpha_{13}}{3 p^2}\left(1152+2384 e^2+509 e^4\right)\Big] \\
& +\frac{\alpha_{52} \left(1-e^2\right)^{3/2}}{15 p^5} \left(36 e^6+1273 e^4-12 \left(7 e^2+8\right) \left(1-e^2\right)^{3/2}+4 e^2-672\right) \\
& +\frac{\epsilon_{3}\left(1-e^2\right)^{3/2}}{15 p^5} \Big[1152+2384 e^2+509e^{4}\Big]\,,
\end{aligned}
\end{equation}
\begin{equation}
\begin{aligned}\label{dedtn}
\Big\langle\frac{de}{dt}\Big\rangle =& -\frac{e \left(1-e^2\right)^{3/2}}{15p^{4}}\Big[121 e^2+304-\frac{a}{2 p^{3/2}} \left(1313 e^4+5592 e^2+7032\right)+\frac{2 \alpha_{13}}{p^2} \left(385 e^4+2664 e^2+2292\right)\Big] \\
& +\frac{\alpha_{52}\left(1-e^2\right)^{3/2}}{60 e p^6} \left(72 e^8+5249 e^6+6562 e^4-2 e^2 \left(121 e^2+304\right) \left(1-e^2\right)^{3/2}-1872 e^2+384\right) \\
& +\frac{\epsilon_{3} e \left(1-e^2\right)^{3/2}}{60 p^6} \Big[9168+10656 e^2+1540 e^4\Big]\,.
\end{aligned}
\end{equation}
The semi-latus rectum ($p$) and the eccentricity ($e$) decrease with time, as long as the contributions coming from the terms proportional to the deformation parameters are subleading, as a result of gravitational radiation since the sign of the leading terms in right sides of Eq.(\ref{dpdtn}) and Eq.(\ref{dedtn}) are negative. Here, one can also easily track the leading PN orders for various correction terms due to the spin and the deformation parameters similar to the flux expressions. Again, we can see by restoring powers of $c,$ the contribution of leading order corrections coming from the deviation parameters emerge at 2PN order.\footnote{ The relative changes in ($p(t), e(t)$) after subtracting out the GR contribution as a function of time have been presented in the Appendix (\ref{apenteu2}) for distinct values of deviation parameters. }

One can further examine how long it takes for the inspiralling object to reach the LSO. It will depend on the relative signs of the deviation parameters and their corresponding LSO values from Eq.(\ref{sp1}). We use Eq.(\ref{dpdtn}) and integrate it from the starting point of the inspiral $p=14$ to $p=p_{sp}$ for determining the time it takes to reach LSO. More precisely, we will consider that change of this time scale, denoted by $\Delta t$, due to the presence of the deviation parameters compared to that of GR. Hence, after integrating Eq.(\ref{dpdtn}) to find the time scale, we subtract the GR part from it. Finally, we get 
\begin{equation}
\begin{aligned}\label{days}
\Delta t \approx & \frac{(1-e^{2})^{-3/2}}{(8+7e^{2})^{2}}\Big[ -\frac{5}{192}(1152+2384e^{2}+509e^{4})(196-p_{sp}^{2})(\alpha_{13}-\frac{\epsilon_{3}}{2})+\\
& \frac{5\alpha_{52}}{384}\Big(36e^{6}-96(7+\sqrt{1-e^{2}}) +4e^{2}(1+3\sqrt{1-e^{2}})  +e^{4}(1273+84\sqrt{1-e^{2}})\Big)(196-p_{sp}^{2})\Big]\,.
\end{aligned}
\end{equation}
where one should only consider the GR part in the expression for $p_{sp}$ defined in Eq.(\ref{sp1}) as we are working only in linear order of the deviation parameters. 
 Note that the  PN corrections coming from the GR part will be cancelled in $\Delta t.$ This then estimates the change of time required by the test body to reach LSO due to the deviation parameters only. 
Thus, these features imply the measurable effects of the Johannsen parameters. Also it is evident from Eq.(\ref{days}), this change of the time scale for the secondary to reach LSO, depends on the relative magnitude of the deviation parameters. Depending on that, if $\Delta t$ becomes negative, then it corresponds to the fact that the secondary takes less time to reach LSO when the deviation parameters are turned on.  Next, to analyze the detectability of deviations in the Kerr black hole, we next examine the orbital phase and GW dephasing.

\section{Detection prospects with GW dephasing}\label{dephasing1}

This section examines the prospects of detecting deviations from the Kerr black hole by considering an EMRI system. As mentioned earlier in Sec. (\ref{gdscvecmtn}), the eccentric motion exhibits two fundamental frequencies ($\Omega_{\phi}, \Omega_{r}$), that is, angular and radial, respectively. In principle, both frequencies have contributions in the phase computation and collectively can be written as
\begin{align}
\frac{d\varphi_{i}}{dt} = \langle \Omega_{i}(p(t),e(t))\rangle = \frac{1}{T_{r}}\int_{0}^{2\pi}d\chi\frac{dt}{d\chi} \Omega_{i}(p(t),e(t),\chi) \hspace{5mm} ; \hspace{5mm} i = (\phi, r),
\end{align}
where $(p(t),e(t))$ will be obtained from Eq.(\ref{dpdtn}) and Eq.(\ref{dedtn}). $\langle \Omega_{i}(p(t),e(t))\rangle$ denotes the orbital averaged frequency. We primarily focus on the azimuthal phase shift regulating the orbital dephasing: $\varphi_{\phi}(t)\sim \phi(t)$ \cite{Barsanti:2022ana, Rahman:2023sof}. We take the initial phase to be zero ($\varphi_{i}(0)=0$). With this, using Eq.(\ref{gdscs3n}), the corresponding expression is given by
\begin{align}\label{orbphase}
\frac{d\phi}{dt} =& \frac{\left(1-e^2\right)^{3/2}}{p^{3/2}}-\frac{a \left(1-e^2\right)^{3/2}}{p^3}\left(3 e^2+1\right)+\frac{\alpha_{52} \left(1-e^2\right)}{4 p^{7/2}} \left(2 e^4-\left(3 \sqrt{1-e^2}+4\right) e^2+2\right) \nonumber\\
&
+\frac{3(1-e^{2})^{3/2}(1+e^{2})}{2p^{7/2}}\Big(\alpha_{13}-\frac{\epsilon_{3}}{2}\Big).
\end{align}

Note that here ($p, e$) are functions of time and are obtained from Eq.(\ref{dpdtn}) and Eq.(\ref{dedtn}) that help us further in solving the Eq.(\ref{orbphase}) to find $\phi(t)$. We use the Mathematica-based built-in function $\mathtt{NDSolve}$ to determine the time evolution of the orbital parameters and phase.  Also note that,  structure of  phase $\phi(t)$ is the following: $\phi(t) \sim \frac{1}{c^{0}}-\frac{a}{c^{3}}+\frac{\alpha_{52}}{c^{4}}+\frac{\alpha_{13}}{c^{4}}-\frac{\epsilon_{3}}{c^{4}}.$ The deviation parameters enter at 2PN order. At this point, note that the complete expression for the phase up to 2PN order will require the addition of the GR contributions at 1PN-2PN order, including the correction due to the spin parameter. However, we will eventually focus on the dephasing, subtracting the GR contribution. Hence, we will not consider these higher-order PN terms for GR. Interested readers are referred to \cite{Moore:2016qxz} for more details.

Let us now analyze the effects of deviation parameters in GW dephasing that will implicate the possible detectability of deformations in Kerr black holes. We use the fact that the GW dephasing is twice the orbital phase, i.e., $\Phi_{\textup{GW}} = 2\phi(t)$. We use an EMRI system with a Johannsen black hole as the reference waveform to study the GW signals for possible deviations from the Kerr black hole. The dephasing between a Johannsen spacetime and a Kerr black hole, up to a given time $t_{\textup{obs}}$, is defined as
\begin{align}
\Delta\Phi(t_{\textup{obs}}) = \vert \Phi^{\textup{JHN}}_{\textup{GW}}(t_{\textup{obs}}) - \Phi^{\textup{Kerr}}_{\textup{GW}}(t_{\textup{obs}}) \vert,
\end{align}
where $\Phi^{\textup{JHN}}_{\textup{GW}}$ represents the GW phase from the Johannsen spacetime and $\Phi^{\textup{Kerr}}_{\textup{GW}}$ represents the GW phase from the Kerr black hole. Their difference provides us with the GW dephasing. We analyze our results taking the observation period one year ($t_{\textup{obs}}=1$ year). As we are considering an EMRI system, we take the primary black hole with the mass $M=10^{6}M_{\odot}$ and the secondary with the mass $\mu=10M_{\odot}$. This sets up the mass ratio $q=10^{-5}$. We consider $p_{\textup{in}}=14$ to be the start of the inspiral with distinct initial eccentricities ($e_{\textup{in}}$). 
\begin{figure}[h!]
	\centering
	\minipage{0.48\textwidth}
	\includegraphics[width=\linewidth]{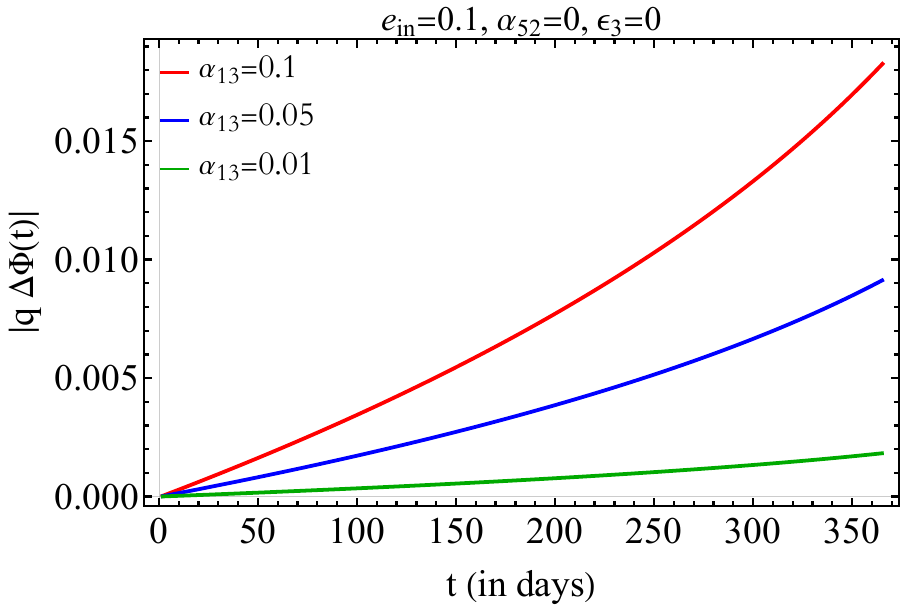}
	\endminipage\hfill
	\minipage{0.48\textwidth}
	\includegraphics[width=\linewidth]{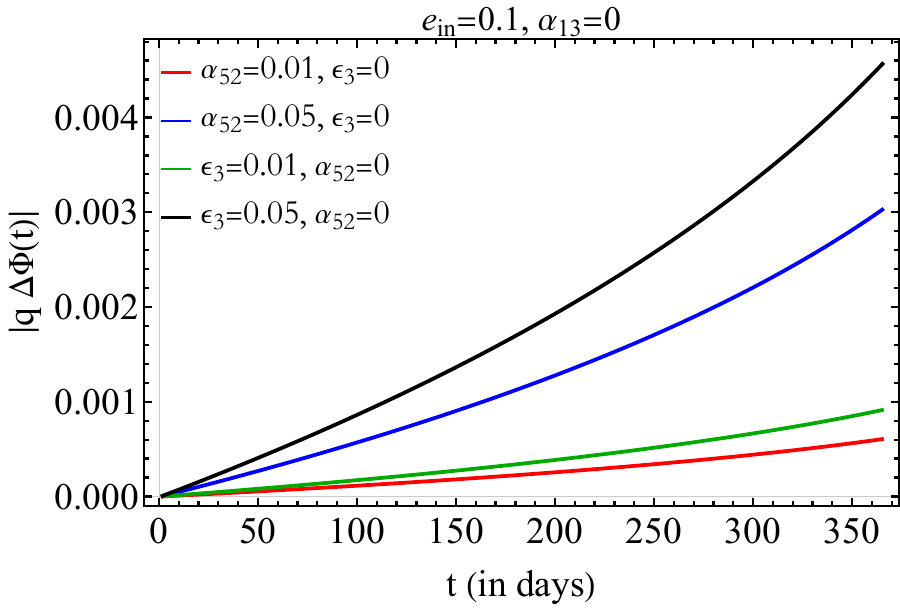}
	\endminipage\hfill
  \minipage{0.48\textwidth}
         \includegraphics[width=\linewidth]{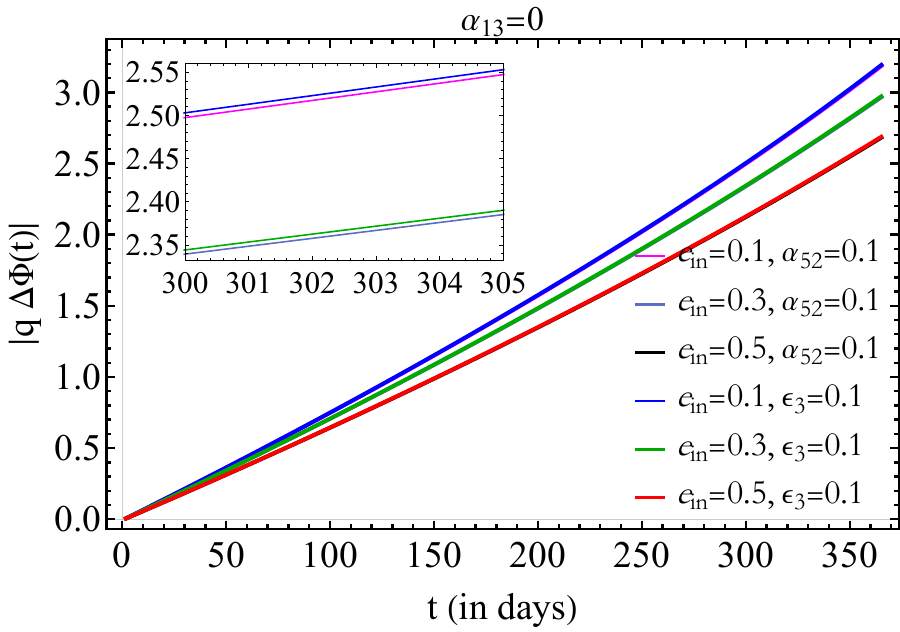}
	 \endminipage\hfill
  \minipage{0.48\textwidth}
  \includegraphics[width=\linewidth]{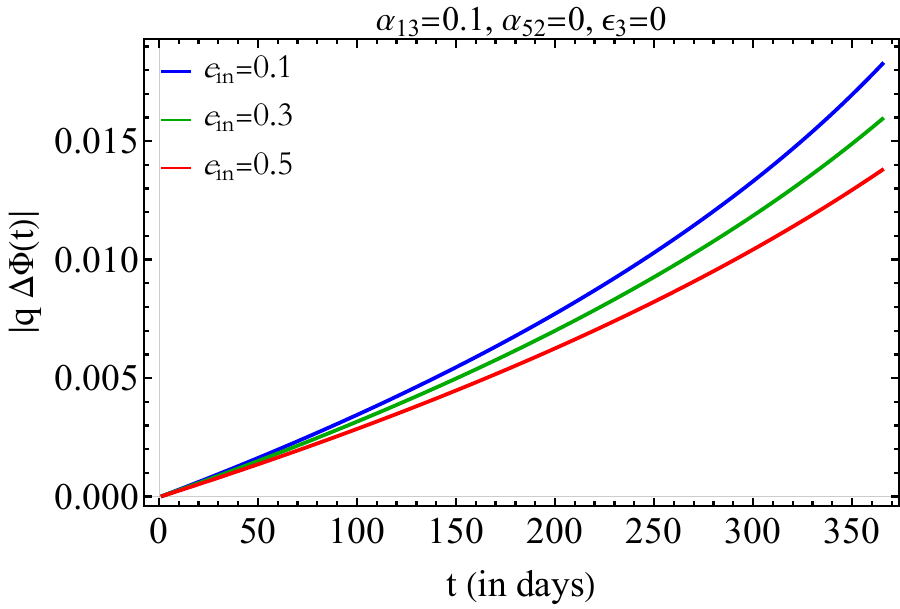}
  \endminipage\hfill
	\caption{The plots represent the dephasing $\Delta\Phi$ in one year of observation time. We consider an EMRI system with the mass of the primary black hole $M=10^{6}M_{\odot}$ and the secondary with the mass $\mu=10M_{\odot}$, setting up the mass ratio $q=10^{-5}$. The upper panel shows the dephasing for different values of ($\alpha_{13}, \alpha_{52}, \epsilon_{3}$) with $e_{\textup{in}} = 0.1$. The lower panel shows the same for different initial eccentricities. 
 }\label{fig_dephase}
\end{figure}

 In Fig.(\ref{fig_dephase})\footnote{One can set $e_\textup{in}=0$ in Eqs. (\ref{dpdtn}, \ref{orbphase}), and obtain the dephasing analytically in the frequency domain: $\Delta\Phi =\frac{25}{16q (M\Omega)^{1/3}}(\alpha_{13}+\alpha_{52}-\epsilon_{3}/2)$. Then, we can provide an order of magnitude estimate of it in the following way. Given the reference \cite{LIGOScientific:2021sio}, we consider $M \sim 140M_{\odot}$, $q \sim 0.60\,.$ And then setting  $\alpha_{13}=0.1$ (and other parameters to be zero) along with the allowed frequency range mentioned in Table (V) of \cite{LIGOScientific:2021sio}, the $\Delta\Phi$ comes out to be in the range of $10^{-2} - 10^{-3}$ which is consistent with the deviation at 2 PN order ($\delta\varphi_{4}) \sim 0.26$ as shown in Table (VI) of \cite{LIGOScientific:2021sio}. Using a similar analysis, we can also show that even when $\alpha_{52}$ and $ \epsilon_{3} \sim 0.1,$ the dephasing is consistent with what comes from LIGO data  \cite{LIGOScientific:2021sio}.} we show the effects of Johannsen parameters on GW dephasing for different initial eccentricities ($e_{\textup{in}}$). Notably, the dephasing eventually depends on the relative signs of the deviation parameters as expressed in Eq. (\ref{orbphase}). The upper panel shows that (for $e_{\textup{in}}=0.1$), the larger the strength of the deviation parameter, the larger the dephasing is. In the lower panel, interestingly, the small eccentricities give rise to a relatively higher magnitude of the dephasing than others. Hence, the dynamics of the inspiralling object, even with lower eccentricity, make the detectability of deviations from the Kerr black hole more promising with LISA. Furthermore, since the secondary object will encounter a strong gravitational field in the vicinity of the last stable orbit, the higher-order PN corrections or the black hole perturbation approach are needed to compute the GW phase more accurately, which will help us more concretely understand this behaviour at late time. Moreover, with an average signal-to-noise ratio (SNR) of about 30 for LISA observations, it should be able to detect the dephasing with $\Delta\Phi \gtrsim 0.1$ rad \cite{PhysRevLett.123.101103}. Note that Johannsen parameters, bringing deviations from the Kerr black hole, cause the dephasing to be proportional to $\mathcal{O}(1/q)$, implying the dependence on the mass ratio and, as a result, giving rise to the notion of detectability from LISA observations.

\begin{figure}[htb!]
	\centering
	\minipage{0.48\textwidth}
	\includegraphics[width=\linewidth]{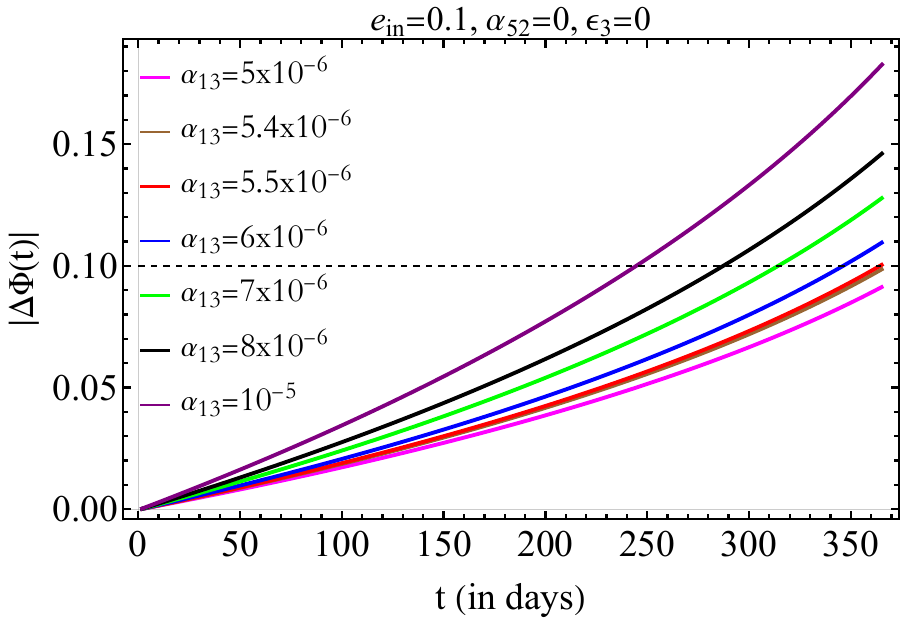}
	\endminipage\hfill
	\minipage{0.48\textwidth}
	\includegraphics[width=\linewidth]{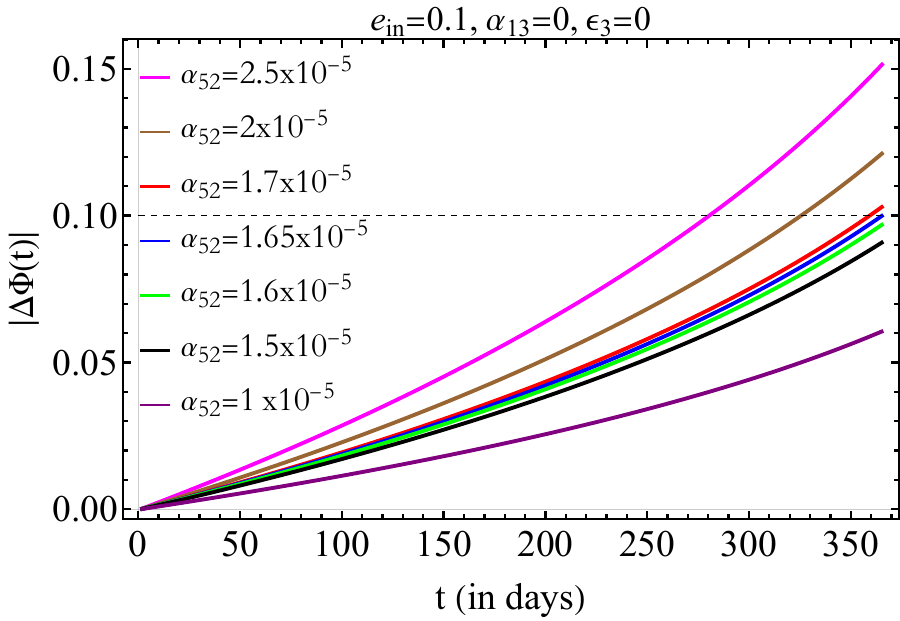}
	\endminipage\hfill
  \minipage{0.48\textwidth}
         \includegraphics[width=\linewidth]{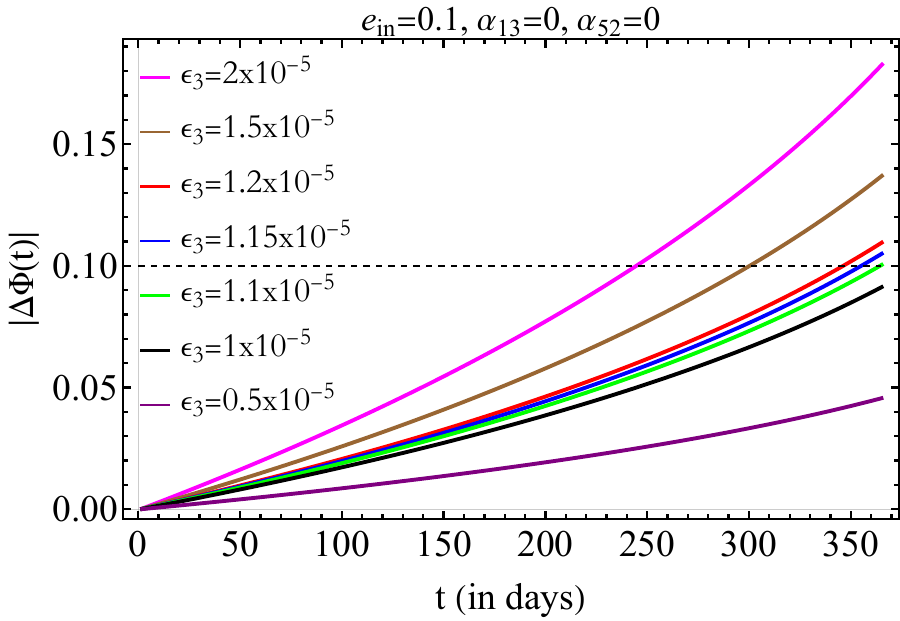}
	 \endminipage
	\caption{The plots provide an order of magnitude estimate of the upper bound on the deviation parameters when $|\Delta\Phi| <0.1\,.$ We set mass ration $q=10^{-5}$ and $e_{\textup{in}}=0.1$. 
 }\label{constrain_dephase}
\end{figure}
{Last but not least, we examine the order of magnitude of the upper bound that these Johannsen parameters will have in case LISA cannot detect dephasing associated with these deviation parameters, i.e. when $|\Delta\Phi| <0.1\,$. This is shown in Fig. (\ref{constrain_dephase}). We consider $q=10^{-5}$ and $e_{\textup{in}}=0.1$, and for $\alpha_{13}\lesssim 5.5 \times 10^{-6}$, $\alpha_{52}\lesssim 10^{-5}$ and $\epsilon_{3}\lesssim 10^{-5}$, the dephasing $|\Delta\Phi(t)|$ remains less than  $0.1$ till the end of the evolution as can be seen from Fig. (\ref{constrain_dephase}).}
This gives an order of magnitude estimate of the upper bound for deviation parameters in case we cannot detect the dephasing using LISA.


\section{Discussion}\label{dscn}

One of the key research objectives of the LISA mission is to map Kerr spacetime \cite{Glampedakis_2006}. To conduct such a major test of GR, we need to be able to set up the required tools that can measure any deviations from the Kerr metric. This will offer deep insights into the weak and strong gravity regimes of the supermassive black hole (SMBH) that can capture a stellar-mass object which emits gravitational radiation while inspiralling the SMBH. The generated GWs carry information of the spacetime geometry, hence acting as a probe for the GR test. In this direction, EMRIs are the perfect astrophysical objects for testing GR with LISA; thus, the comprehensive mapping of the spacetime metric, where the inspiralling object moves, should be made possible by the detection of GWs from EMRIs \cite{Glampedakis_2005}. Here, we attempt to analyze such a notion, which brings the deviations in Kerr metric and can have possible detectability through low-frequency detectors.

We consider the Johannsen spacetime that carries four parameters ($\alpha_{13}, \alpha_{52}, \epsilon_{3}, \alpha_{22}$) which infer the deviations from the Kerr black hole \cite{PhysRevD.88.044002, Staelens:2023jgr}. The broader scope of deviations is somewhat restricted by the perturbative method. We limit the deformation parameters in the metric to small deviations from the GR solution. Hence, we examine our results with leading order PN corrections. We determine the eccentric equatorial orbital motion of the inspiralling object and find the expression for the last stable orbit, i.e., the truncation region of the inspiralling object, which complies with \cite{Glampedakis:2002ya}. We perform our analysis within the adiabatic approximation that breaks down beyond the LSO; therefore, we confine the motion in the domain ($p_{\textup{in}}=14, p_{\textup{min}}(e)$). We analytically calculate instantaneous fluxes and compute GW fluxes. With the average loss of energy and angular momentum fluxes, we obtain the expressions for orbital evolution parameters ($p(t), e(t)$) and estimate how long it takes for the inspiralling object to reach the LSO. We also analyse the relative change in eccentric orbital parameters due to the effect of deformations and find that the relative change in ($p(t), e(t)$) increases with a larger magnitude of deviations.  Note that all results with the Kerr limit in slow-rotation approximation are in agreement with \cite{Flanagan:2007tv} if considered non-equatorial orbits.


Next, we estimate GW dephasing and further analyze the detection prospects of deviations from the Kerr black hole with the mass ratio $q=10^{-5}$. We find that the changes in deviation parameters, shown in Fig. (\ref{fig_dephase}), cause dephasing to be prominent from the detection perspective as it goes as $\mathcal{O}(1/q)$.  We observe that the leading order corrections from the deviation parameters appear at 2PN, whereas spin contributes at 1.5PN. We have ignored the general relativistic 1PN-2PN contributions \footnote{However, readers are suggested to refer \cite{Moore:2016qxz} for comprehensive details of GR contributions up to 3PN order.} in the calculations as our main goal was to study the dephasing. The PN terms of GR contributing to final results would be eliminated as we are computing the phase difference between the Johannsen and Kerr case. It turns out that $\alpha_{22}$ does not play any role to the results derived in the leading order PN corrections; however, other parameters ($a, \alpha_{13}, \alpha_{52}, \epsilon_{3}$) provide measurable effects. We further notice that $\epsilon_{3}$ generates a relatively larger dephasing than the other parameters present in the metric. The larger the deviation parameters are, the more dephasing there is. Further, the dynamics with lower initial eccentricities produce larger dephasing. Since we know that observations over the year with the average SNR 30, LISA should detect dephasing $\Delta\Phi \gtrsim 0.1$ rad \cite{PhysRevLett.123.101103}; thus, our analysis provides measurable effects with such low-frequency detectors. We have also provided an estimate for the upper bound of these deviation parameters in case LISA can not detect the associated dephasing.

However, we strongly expect that the higher-order PN corrections will be an interesting addition to such an analysis, which we hope to communicate in our upcoming studies. Another direction to explore these scenarios is the black hole perturbation \cite{Li:2022pcy, Pound:2021qin} that can more concretely improvise the PN results and can constrain the deviation parameters with their possible detectability. If LISA might impose interesting limits on non-GR deviations/deformations, it is an intriguing topic that could be addressed with the aid of the Johannsen framework. In order to execute such an analysis, one needs to evolve geodesics in this background and develop the corresponding GWs.  To quantify more precisely the magnitude of the non-GR deviations from the Kerr black hole using  LISA observations, one might next perform a Fisher type of analysis \cite{Vallisneri:2007ev}. These are some of the aspects that will advance the understanding of deviations in Kerr, which we would like to explore in future studies.

\section*{Acknowledgements} 
S. K. would like to thank Mostafizur Rahman for useful discussions. A.B. would like to thank the speakers of the workshop ``Testing Aspects of  General Relativity-II" (11-13th April 2023) and ``New insights into particle physics from quantum information and gravitational waves" (12-13th June 2023) at Lethbridge University, Canada, funded by McDonald Research Partnership-Building Workshop grant by McDonald Institute for useful discussions. Research of S. K. is supported by the research grant (202011BRE03RP06633-BRNS) by the Board of Research In Nuclear Sciences (BRNS), Department of Atomic Energy (DAE), India. A. C. is supported by the Prime Minister's Research Fellowship (PMRF-192002-1174) of the Government of India. A.B. is supported by the Mathematical Research Impact Centric Support Grant (MTR/2021/000490) by the Department of Science and Technology Science and Engineering Research Board (India) and the Relevant Research Project grant (202011BRE03RP06633-BRNS) by the Board Of Research In Nuclear Sciences (BRNS), Department of Atomic Energy (DAE), India. A.B. also acknowledge associateship program of Indian Academy of Science, Bengaluru. Authors would like to thank the anonymous referee for useful comments and suggestions.
  

\appendix

\section{Geodesic velocities and constants of motion} \label{apenteu1}
In this section, we derive velocity components of geodesic motion exhibited by the inspiralling object in the Johannsen background. Here, we obtain the equations without setting the equatorial assumption. We implement the Hamilton-Jacobi formalism; the action for which is given as \cite{Johannsen:2013szh, Staelens:2023jgr}
\begin{align}\label{ac1}
S = \frac{1}{2}\mu^{2}\tau-Et+J_{z}\phi+R(r)+\Theta(\theta),
\end{align}
 and we have
\begin{align}
-\frac{\partial S}{\partial\tau} = \frac{1}{2}g^{\alpha\beta}\frac{\partial S}{\partial x^{\alpha}}\frac{\partial S}{\partial x^{\beta}}.
\end{align}
with $p_{\alpha}=\partial S/\partial x^{\alpha}$ and $\partial S/\partial\tau = \mu^{2}/2$. The separable equations in radial and angular terms take the following form
\begin{align}\label{j1}
\Delta A_{5} \Big(\frac{dR}{dr}\Big)^{2}-\frac{1}{\Delta}\Big(A_{1}(r^{2}+a^{2})E-aA_{2}J_{z}\Big)^{2}+\mu^{2}\Big(r^{2}+\epsilon_{3}\frac{M^{3}}{r}\Big) = -Q, \\
\Big(\frac{d\Theta}{d\theta}\Big)^{2}+a^{2}\mu^{2}\cos^{2}\theta+\Big(aE\sin\theta-\frac{J_{z}}{\sin\theta}\Big)^{2} = Q. \label{j2}
\end{align}
Carter constant is given as: $\mathcal{Q}\equiv Q-(J_{z}-aE)^{2}$. The functions $R(r)$ and $\Theta(\theta)$ can further be related to the respective momenta of the particle,
\begin{align}\label{j3}
\frac{dR}{dr} = \frac{\mu\Sigma}{\Delta A_{5}}\frac{dr}{d\tau} \hspace{7mm} ; \hspace{7mm} \frac{d\Theta}{d\theta} = \mu \Sigma \frac{d\theta}{d\tau}\,.
\end{align}
Thus, the Eq.(\ref{j1}) and Eq.(\ref{j2}) can be written as
\begin{align}
\mu^{2}\Big(\frac{dr}{d\tau}\Big)^{2} =& \frac{A_{5}}{\Sigma^{2}}\Big[\Big(A_{1}(r^{2}+a^{2})E-aA_{2}J_{z}\Big)^{2}-\Delta\Big(Q+\mu^{2}r^{2}+\mu^{2}\epsilon\frac{M^{3}}{r}\Big)\Big], \\
\mu^{2}\Big(\frac{d\theta}{d\tau}\Big)^{2} =& \frac{1}{\Sigma^{2}}\Big[(Q-a^{2}\mu^{2}\cos^{2}\theta)-\Big(aE\sin\theta-\frac{J_{z}}{\sin\theta}\Big)^{2}\Big].
\end{align}
Since we have two conserved quantities- energy and angular momentum, we can obtain the following relations,
\begin{align}\label{gdscs}
\mu\frac{dt}{d\tau}  =&\frac{1}{\mathcal{C}}\Big[N \left(A_{1} \left(a^2+r^2\right) \left(a^2 A_{1} E-a A_{2} J_{z}+A_{1} E r^2\right)-a^2 \Delta E \sin ^2(\theta )+a \Delta  J_{z}\right)\Big]\,, \\
\mu\frac{d\phi}{d\tau} =& \frac{1}{\mathcal{C}}\Big[N \left(a A_{2} \left(a^2 A_{1} E-a A_{2} J_{z}+A_{1} E r^2\right)-a \Delta  E+\Delta  J_{z} \csc ^2\theta \right)\Big],
\end{align}
where $\mathcal{C}=\Delta  \Sigma  \left(A_{1} \left(a^2+r^2\right)-a^2 A_{2} \sin ^2(\theta )\right)^2$. The expressions in the slow-rotation approximation become,

\begin{align}\label{gdscs2}
\mu \frac{dt}{d\tau} =& \Big[-(r^{2}A_{1}A_{2}-\Delta)\frac{aJ_{z}}{r^{2}\Delta}+\frac{Er^{2}}{\Delta}A_{1}^{2}\Big]\Big(1+\epsilon_{3}\frac{M^{3}}{r^{3}}\Big)^{-1} + \mathcal{O}(a^{2})\,,\\ \label{gdscs3}
\mu \frac{d\phi}{d\tau} =& \Big[\frac{J_{z}}{r^{2}\sin^{2}\theta}+\frac{aE}{r^{2}\Delta}(A_{1}A_{2}r^{2}-\Delta)\Big]\Big(1+\epsilon_{3}\frac{M^{3}}{r^{3}}\Big)^{-1}+ \mathcal{O}(a^{2})\,, \\ \label{gdscs4}
\mu^{2}\Big(\frac{dr}{d\tau}\Big)^{2} =& \frac{A_{5}}{\Sigma^{2}}\Big[(A_{1}r^{2}E-aA_{2}J_{z})^{2}-\Delta\Big(Q+\mu^{2}r^{2}+\mu^{2}\epsilon_{3}\frac{M^{3}}{r}\Big)\Big] + \mathcal{O}(a^{2})\,, \\ \label{gdscs5}
\mu^{2}\Big(\frac{d\theta}{d\tau}\Big)^{2} =& \frac{1}{\Sigma^{2}}\Big[Q-\Big(aE\sin\theta-\frac{J_{z}}{\sin\theta}\Big)^{2}\Big] + \mathcal{O}(a^{2}).
\end{align}
In the linear-order approximation of $a$, $\Sigma=r^{2}+\epsilon_{3} \frac{M^{3}}{r}$ and $\Delta = r^{2}-2Mr$. Using the four velocities,
\begin{align}
\mu^{2}\Big[\Big(\frac{dr}{d\tau}\Big)^{2}+r^{2}\Big(\frac{d\theta}{d\tau}\Big)^{2}+r^{2}\sin^{2}\theta\Big(\frac{d\phi}{d\tau}\Big)^{2} \Big] =& 
(E^2-\mu ^2)\Big(1+\frac{\alpha_{52}M^{2}}{r^{2}}\Big)-\frac{M^{2}(2 \mu ^2 M^2 \epsilon_{3}+\alpha_{52} Q)}{r^4} +\frac{2 \mu ^2 M}{r} \nonumber \\ &+\frac{M^{3}\Big(2 \alpha_{13} E^2+\epsilon_{3} \left(\mu ^2 -2 E^2 \right)+2 \alpha_{52} \mu ^2\Big)+2 M Q}{r^3}\,.
\end{align}
Here, we have ignored terms with $\mathcal{O}\Big(\frac{aM^{2}}{r^{4}}\Big)$ and $\mathcal{O}\Big(\frac{aM}{r^{3}}\Big)$. Further, omitting $\mathcal{O}(r^{-3})$, the above expression with respect to coordinate time can be written in the following form
\begin{align}
\mu^{2}\Big[\Big(\frac{dr}{dt}\Big)^{2}+r^{2}\Big(\frac{d\theta}{dt}\Big)^{2}+r^{2}\sin^{2}\theta\Big(\frac{d\phi}{dt}\Big)^{2} \Big] = &  \frac{\mu^{2}}{E^{2}}\Big[E^{2}-\mu^{2}+\frac{2M}{r}(3\mu^{2}-2E^{2}) \nonumber \\
& +\frac{M^{2}}{r^{2}} \Big(E^{2}(4+\alpha_{52})-(12+\alpha_{52})\mu^{2}\Big)\Big]\,.
\end{align}
Replacing $E=\mu+\mathcal{E}$ \cite{Misner:1973prb, Ryan:1995xi, PhysRevD.78.064028} for separating out the rest mass energy and considering terms linear order in $\mathcal{E}$ and ignoring $\mathcal{O}(\mathcal{E} M/r)$ with their higher order terms. Following \cite{Mukherjee:2022pwd}, we define
\begin{align}\label{ener2}
\mathcal{E} = \frac{\mu}{2}\Big[\Big(\frac{dr}{dt}\Big)^{2}+r^{2}\Big(\frac{d\theta}{dt}\Big)^{2}+r^{2}\sin^{2}\theta\Big(\frac{d\phi}{dt}\Big)^{2} \Big]-\frac{\mu M}{r}\,.
\end{align}

With this, the equatorial plane reduces the velocities in the following form\footnote{Since we are focusing only on leading order PN corrections in deviation parameters that arise at 2PN; as a result, we can discard term $\frac{2J_{z}^{2}}{r^{2}}$ appearing with $\epsilon_{3}$.}
\begin{equation}
\begin{aligned}\label{gdscs3n}
\Big(\frac{dr}{d\tau}\Big)^{2} = & 2 \mathcal{E}+\frac{2}{r}-\frac{J_{z}^2}{r^2}-\frac{4 a J_{z}}{r^3}+\frac{2 \alpha_{13}}{r^3}-\frac{\epsilon_{3}}{r^{3}} \left(1-\frac{2 J_{z}^2}{r^2}\right)+\frac{\alpha_{52}}{r^{3}} \left(2-\frac{J_{z}^2}{r}\right)\,, \\
\frac{d\phi}{d\tau} =& \frac{J_{z}}{r^2}+\frac{2 a }{r^3}-\frac{\epsilon_{3} J_{z} }{r^5}\,.
\end{aligned}
\end{equation}
These velocities will be useful while computing the fluxes in the main text. However, for higher order corrections and more precise results, one should numerically integrate geodesic equations. Further, we use Eqs.(\ref{gdscs3}, \ref{gdscs5}) and compute, 
\begin{align}
\mu^{2}r^{4}\Big[\Big(\frac{d\theta}{d\tau}\Big)^{2}+\sin^{2}\theta\Big(\frac{d\phi}{d\tau}\Big)^{2} \Big] = Q+2 a E J_{z}\Big(1+\frac{2 M}{r}+\frac{M^2\alpha_{22}}{r^2}\Big)\,.\end{align}
We notice that the right-hand side is linear in $a$ and $Q$. We can re-write the above expression coordinate time $t$,
\begin{align}
\mu^{2}r^{4}\Big[\Big(\frac{d\theta}{dt}\Big)^{2}+\sin^{2}\theta\Big(\frac{d\phi}{dt}\Big)^{2} \Big] = Q+2 a E J_{z}+\frac{4 a E J_{z} M}{r}+\frac{2 a E J_{z} M^2\alpha_{22}}{r^2}\,. \label{ef}
\end{align}
By introducing $\mathcal{Q}$ and defining, $\mathcal{Q}+J_{z}^{2}\equiv Q+2aEJ_{z}$, we may further express the conserved quantity angular momentum,
\begin{align}\label{dphidt}
J_{z} = \mu r^{2}\sin^{2}\theta \Big(1+\epsilon_{3}\frac{M^{3}}{r^{3}}\Big) \frac{d\phi}{d\tau} -\frac{2 a E M \sin^2 \theta}{r}\,.
\end{align}
The constants of motion are consistent with \cite{PhysRevD.78.064028, Ryan:1995xi}. Eq.(\ref{ef}) to $\mathcal{O}(a)$ takes the following form,
\begin{align}
\mu^{2}r^{4}\Big[\Big(\frac{d\theta}{dt}\Big)^{2}+\sin^{2}\theta\Big(\frac{d\phi}{dt}\Big)^{2} \Big] = \mathcal{Q}+J_{z}^{2}+a\Big(\frac{4EM}{r}+\frac{2EM^{2}\alpha_{22}}{r^{2}}\Big) \Big(1+\epsilon_{3}\frac{M^{3}}{r^{3}}\Big)\mu r^{2}\sin^{2}\theta \frac{d\phi}{d\tau},
\end{align}
or in the linear order corrections
\begin{align}\label{cart1}
\mathcal{Q}+J_{z}^{2} = \mu^{2}r^{4}(\Dot{\theta}^{2}+\sin^{2}\theta\Dot{\phi}^{2})-4a\mu^{2} Mr\Dot{\phi}\sin^{2}\theta.
\end{align}
Since our analysis takes the equatorial consideration, the involvement of the Carter constant is not required. These constants of motion help in estimating the fluxes, and we consider mentioning them in the main text as well.

\section{Orbital evolution}\label{apenteu2}
Here, we analyze the relative change in eccentric orbital parameters ($p(t)$, $e(t)$) over time, sourced by the deviation parameter. We define the following: $$\Delta p(t)=\Big\vert p^{\textup{JHN}}(t) - p^{\textup{Kerr}}(t) \Big\vert,\quad  \Delta e(t)=\Big|e^{\textup{JHN}}(t) - e^{\textup{Kerr}}(t) \Big\vert\,.$$ This is obtained by subtracting the PN contributions of the GR part from Eqs. (\ref{dpdtn}, \ref{dedtn}). These reflect how the orbital dynamics change due to deviation parameters alone (which enter at 2PN order as discussed in the main text) as a function of time.  

\begin{figure}[htb!]
	\centering
	\minipage{0.33\textwidth}
	\includegraphics[width=\linewidth]{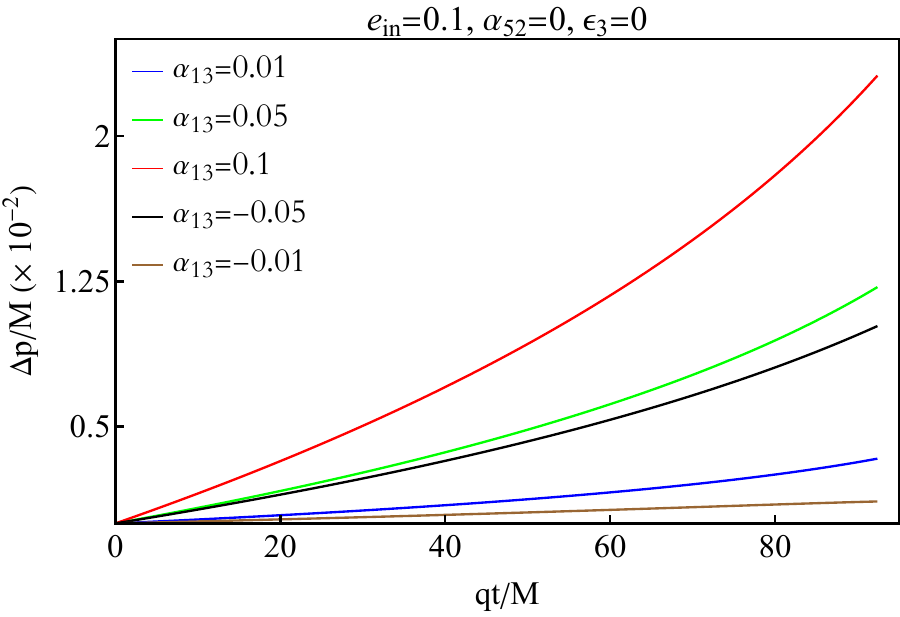}
	\endminipage\hfill
	\minipage{0.33\textwidth}
	\includegraphics[width=\linewidth]{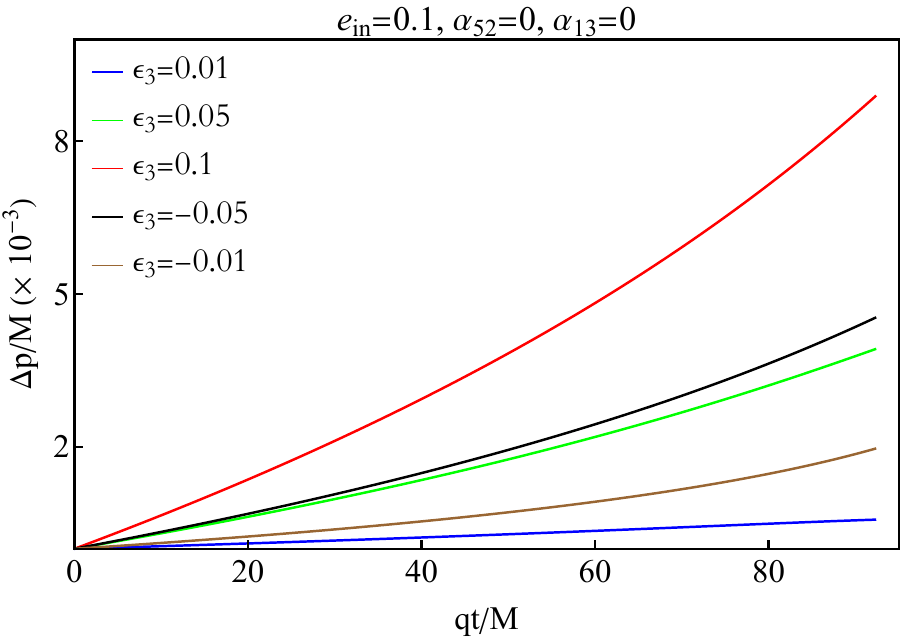}
	\endminipage\hfill
  \minipage{0.33\textwidth}
         \includegraphics[width=\linewidth]{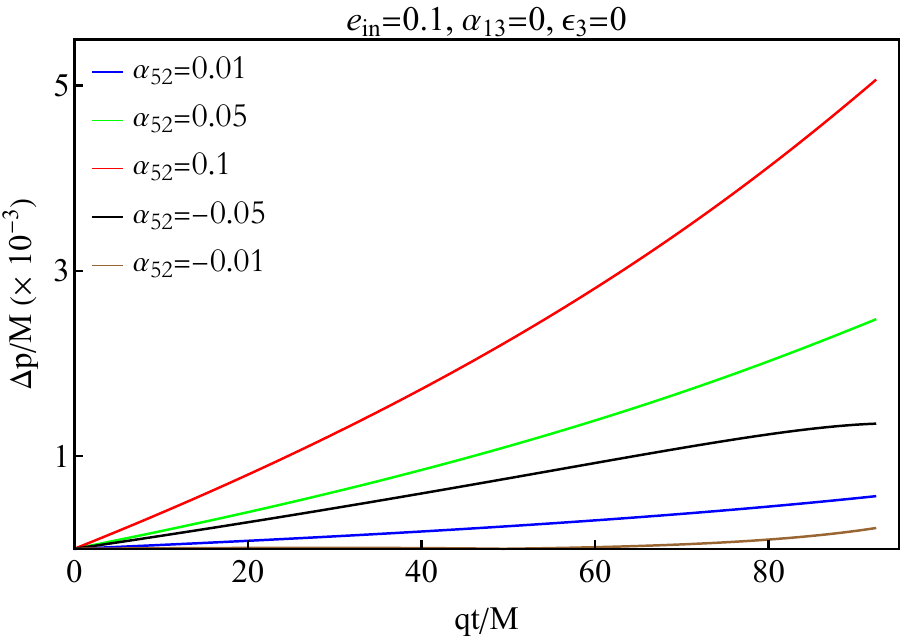}
	\endminipage\hfill
 \minipage{0.33\textwidth}
	\includegraphics[width=\linewidth]{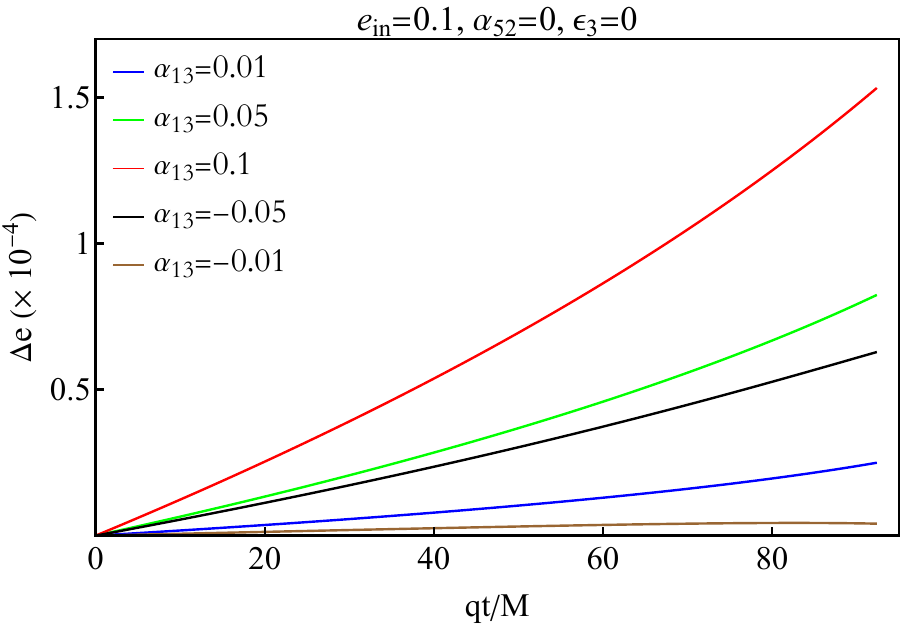}
	\endminipage\hfill
  \minipage{0.33\textwidth}
         \includegraphics[width=\linewidth]{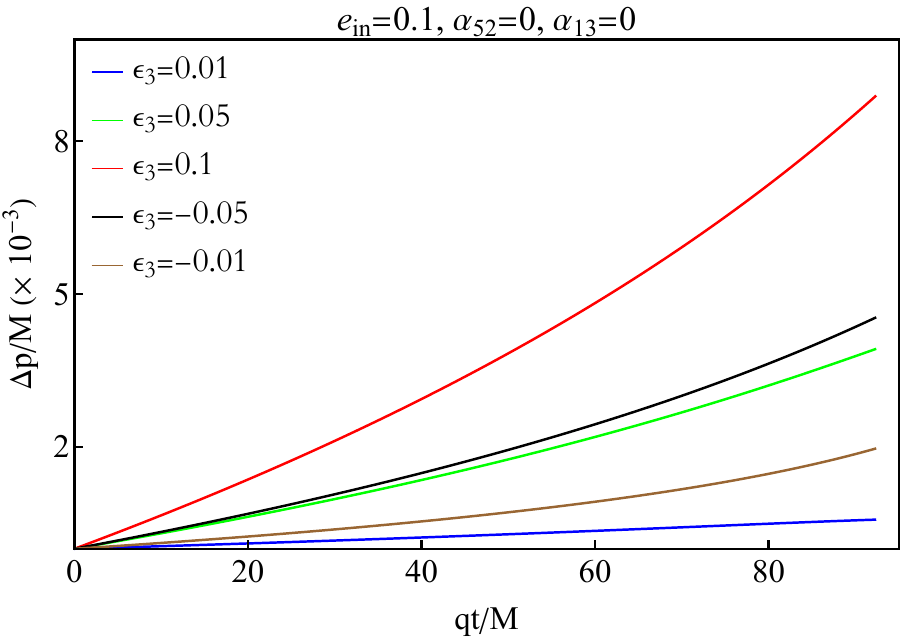}
	 \endminipage\hfill
  \centering
	\minipage{0.33\textwidth}
	\includegraphics[width=\linewidth]{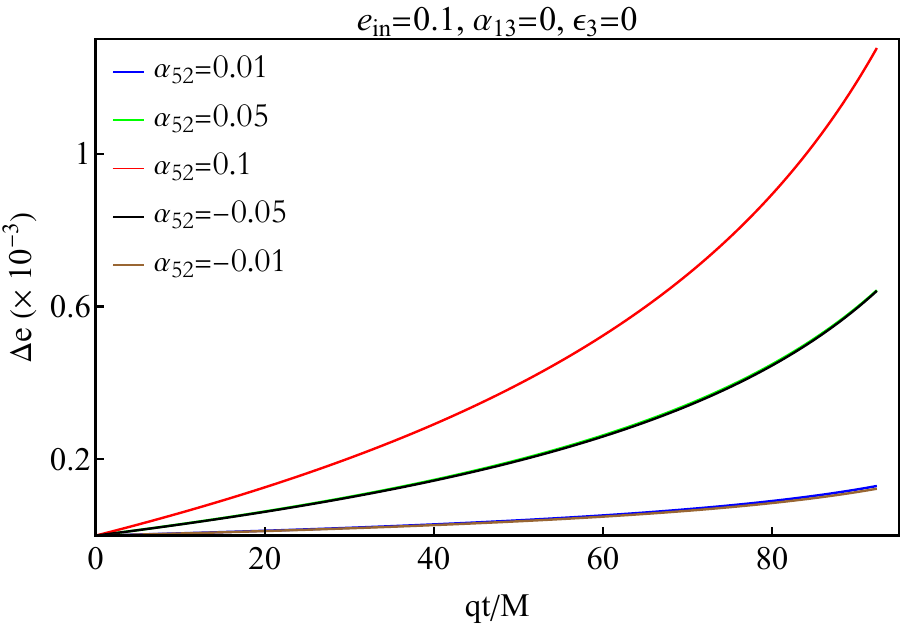}
	\endminipage
	\caption{The upper panel shows the time evolution of relative change in $p$ caused by distinct values of deviation parameters with $e_{\textup{in}}$=0.1. The lower panel provides the time evolution of relative change in orbital eccentricity with a given $e_{\textup{in}}$=0.1. 
 }\label{New_fig_enr_flux_same_e}
\end{figure}
\begin{figure}[htb!]
	\centering
	\minipage{0.33\textwidth}
	\includegraphics[width=\linewidth]{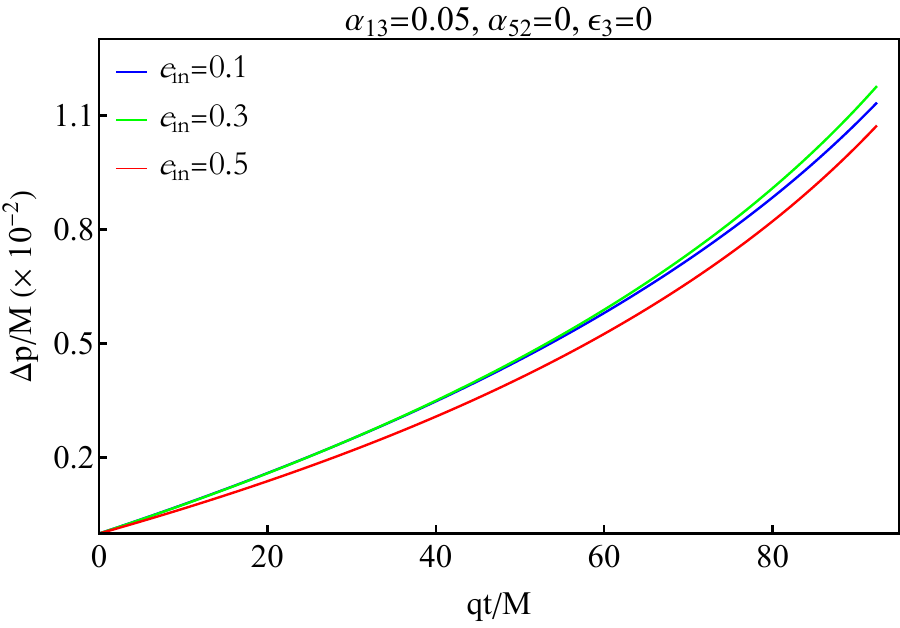}
	\endminipage\hfill
	\minipage{0.33\textwidth}
	\includegraphics[width=\linewidth]{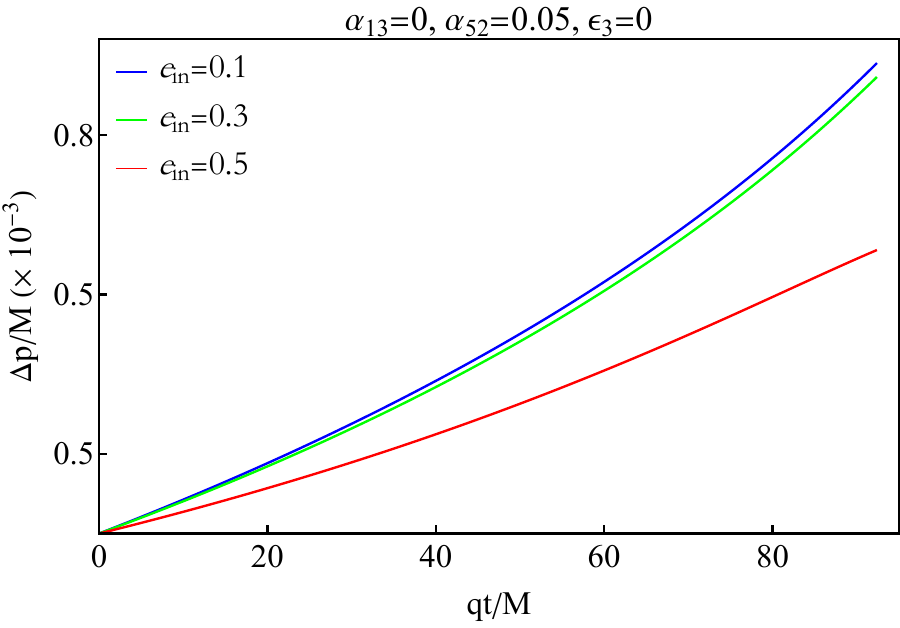}
	\endminipage\hfill
  \minipage{0.33\textwidth}
         \includegraphics[width=\linewidth]{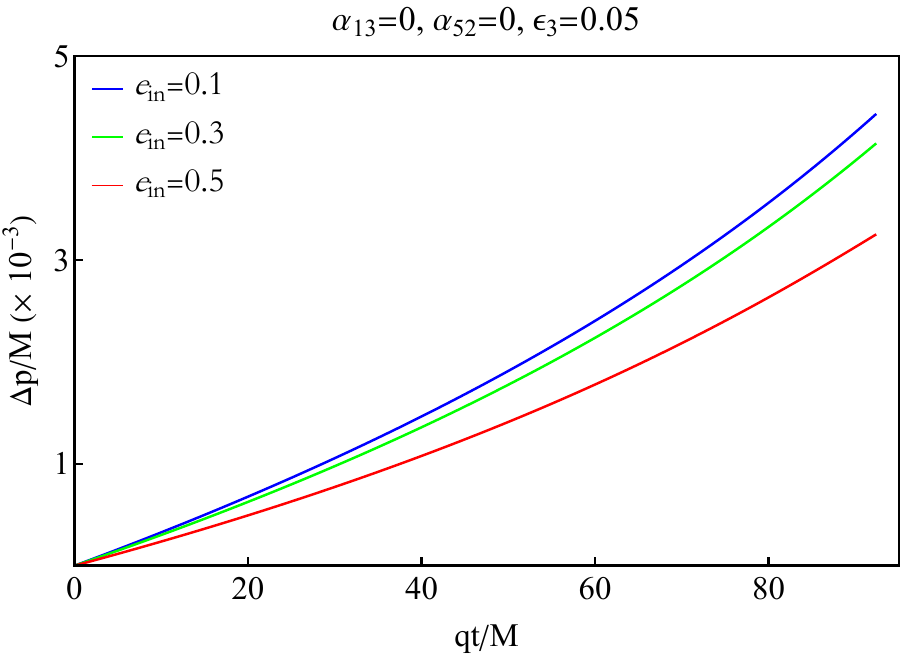}
	\endminipage\hfill
 \minipage{0.33\textwidth}
	\includegraphics[width=\linewidth]{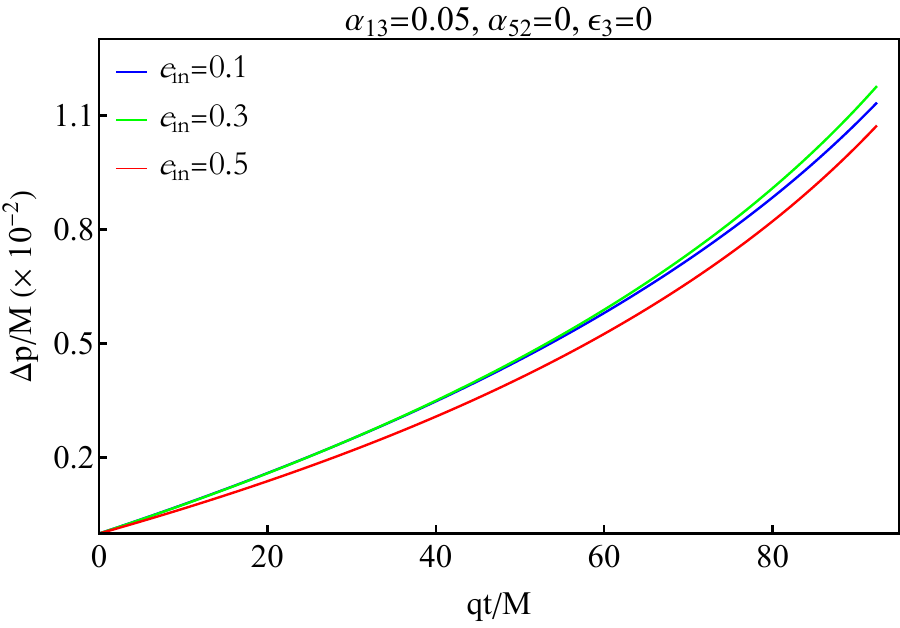}
	\endminipage\hfill
  \minipage{0.33\textwidth}
         \includegraphics[width=\linewidth]{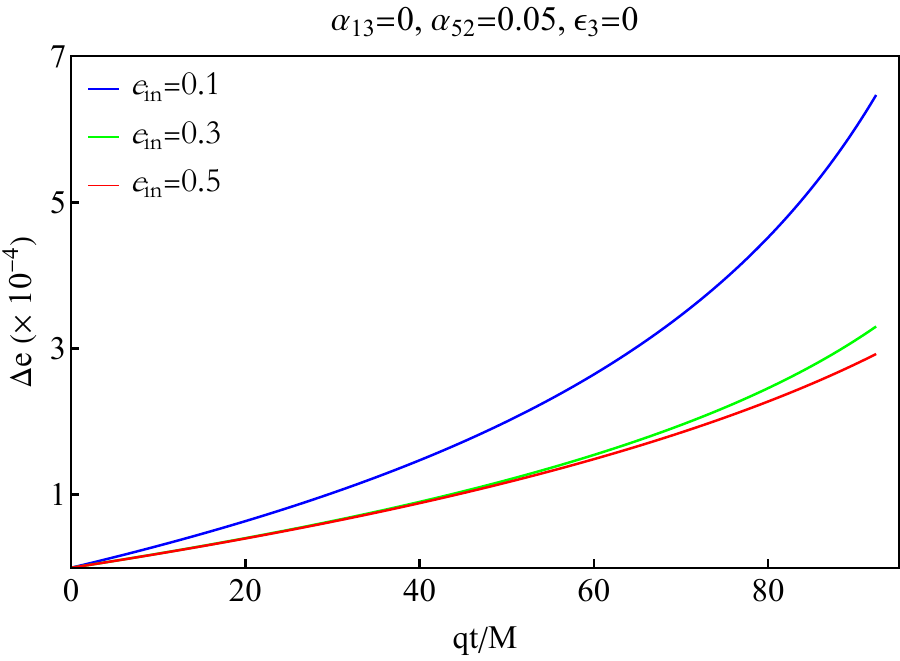}
	 \endminipage\hfill
  \centering
	\minipage{0.33\textwidth}
	\includegraphics[width=\linewidth]{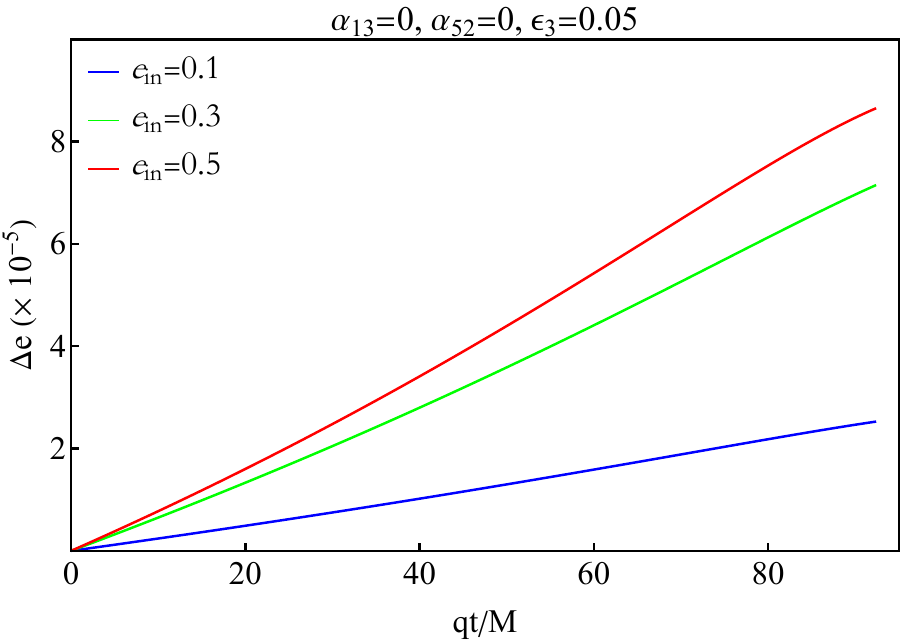}
	\endminipage
	\caption{The time evolution of relative change is $p$ (upper panel), and $e$ (lower panel) is shown for deviation parameters ($\alpha_{13}, \epsilon_{3}, \alpha_{52}$) with distinct initial eccentricities.
 }\label{New_fig_enr_flux_same_e1}
\end{figure}
\vspace{1cm}
Fig. (\ref{New_fig_enr_flux_same_e}) and Fig. (\ref{New_fig_enr_flux_same_e1}) provide an estimate for the relative changes ($\Delta p(t), \Delta e(t)$).  It is evident from the plots that these changes exhibit increasing behaviour over time; however, the order of magnitude depends on values of deviation parameters, which we have depicted in the figures.

\newpage

\bibliography{JN.bib}
\bibliographystyle{JHEP}
\end{document}